%% file: main.tex
\documentclass[ALICE,manyauthors]{cernphprep}
\usepackage[comma,square,numbers,sort&compress]{natbib}
\usepackage{hyperref}
\usepackage{lineno}
\usepackage{xspace}
\usepackage{upgreek}
\usepackage{xcolor}
\usepackage[T1]{fontenc}
\usepackage{orcidlink}
\begin{document}
\input{commands}

\begin{titlepage}
\PHyear{2026}       
\PHnumber{267}      
\PHdate{18 September}  

\title{Measurement of prompt and non-prompt $\pmb{\Dstar}$-meson spin alignment in pp collisions at $\pmb{\s=13.6~\TeV}$}
\ShortTitle{Prompt and non-prompt \Dstar spin alignment in pp at $\s=13.6~\TeV$}   

\Collaboration{ALICE Collaboration\thanks{See Appendix~\ref{app:collab} for the list of collaboration members}}
\ShortAuthor{ALICE Collaboration} 

\begin{abstract}

A precise measurement of prompt and non-prompt \Dstar-meson spin alignment with respect to the helicity and production axes in proton-proton (pp) collisions at $\sqrt{s}~=~13.6$~Te\kern-.1emV\xspace is presented, using data collected with the ALICE experiment at the LHC. The spin alignment is quantified by measuring the diagonal spin density matrix element \rzz for \Dstar mesons at midrapidity ($|y|<0.8$). The measurement is performed in the transverse-momentum (\pt) range $3<\pt<100~\GeVc$ for prompt \Dstar mesons and $3<\pt<30~\GeVc$ for non-prompt \Dstar mesons, respectively. The measured \rzz values for prompt \Dstar mesons are consistent with the unpolarised expectation of $1/3$ over the entire \pt range, indicating no evidence of spin alignment. In contrast, \rzz of non-prompt \Dstar mesons exhibit a deviation from $1/3$, consistent with expectations from weak beauty-hadron decays and well described by PYTHIA~8 simulations coupled with the EvtGen decay package. These results confirm and extend previous measurements, providing a precise baseline for future studies of \Dstar spin alignment in heavy-ion collisions.

\end{abstract}
\end{titlepage}

\setcounter{page}{2} 


\input{intro}
\input{dataset}
\input{analysis}
\input{systematics}
\input{results}


\newenvironment{acknowledgement}{\relax}{\relax}
\begin{acknowledgement}
\section*{Acknowledgements}
\input{fa_2026-08-11_Opt_C.tex}
\end{acknowledgement}

\bibliographystyle{utphys}   
\bibliography{bibliography}

\newpage
\appendix

%
%

\section{The ALICE Collaboration}
\label{app:collab}
\input{Alice_Authorlist_2026-08-11_Opt_C.tex}  
\end{document}

%% file: commands.tex
%

\newcommand{\pp}           {pp\xspace}
\newcommand{\ppbar}        {\mbox{$\mathrm {p\overline{p}}$}\xspace}
\newcommand{\XeXe}         {\mbox{Xe--Xe}\xspace}
\newcommand{\PbPb}         {\mbox{Pb--Pb}\xspace}
\newcommand{\pA}           {\mbox{pA}\xspace}
\newcommand{\pPb}          {\mbox{p--Pb}\xspace}
\newcommand{\AuAu}         {\mbox{Au--Au}\xspace}
\newcommand{\dAu}          {\mbox{d--Au}\xspace}

\newcommand{\s}            {\ensuremath{\sqrt{s}}\xspace}
\newcommand{\snn}          {\ensuremath{\sqrt{s_{\mathrm{NN}}}}\xspace}
\newcommand{\pt}           {\ensuremath{p_{\rm T}}\xspace}
\newcommand{\meanpt}       {$\langle p_{\mathrm{T}}\rangle$\xspace}
\newcommand{\ycms}         {\ensuremath{y_{\rm CMS}}\xspace}
\newcommand{\ylab}         {\ensuremath{y_{\rm lab}}\xspace}
\newcommand{\etarange}[1]  {\mbox{$\left | \eta \right |~<~#1$}}
\newcommand{\yrange}[1]    {\mbox{$\left | y \right |~<~#1$}}
\newcommand{\dndy}         {\ensuremath{\mathrm{d}N_\mathrm{ch}/\mathrm{d}y}\xspace}
\newcommand{\dndeta}       {\ensuremath{\mathrm{d}N_\mathrm{ch}/\mathrm{d}\eta}\xspace}
\newcommand{\avdndeta}     {\ensuremath{\langle\dndeta\rangle}\xspace}
\newcommand{\dNdy}         {\ensuremath{\mathrm{d}N_\mathrm{ch}/\mathrm{d}y}\xspace}
\newcommand{\Npart}        {\ensuremath{N_\mathrm{part}}\xspace}
\newcommand{\Ncoll}        {\ensuremath{N_\mathrm{coll}}\xspace}
\newcommand{\dEdx}         {\ensuremath{\textrm{d}E/\textrm{d}x}\xspace}
\newcommand{\RpPb}         {\ensuremath{R_{\rm pPb}}\xspace}

\newcommand{\nineH}        {$\sqrt{s}~=~0.9$~Te\kern-.1emV\xspace}
\newcommand{\seven}        {$\sqrt{s}~=~7$~Te\kern-.1emV\xspace}
\newcommand{\twoH}         {$\sqrt{s}~=~0.2$~Te\kern-.1emV\xspace}
\newcommand{\twosevensix}  {$\sqrt{s}~=~2.76$~Te\kern-.1emV\xspace}
\newcommand{\five}         {$\sqrt{s}~=~5.02$~Te\kern-.1emV\xspace}
\newcommand{\twosevensixnn}{$\sqrt{s_{\mathrm{NN}}}~=~2.76$~Te\kern-.1emV\xspace}
\newcommand{\fivenn}       {$\sqrt{s_{\mathrm{NN}}}~=~5.02$~Te\kern-.1emV\xspace}
\newcommand{\LT}           {L{\'e}vy-Tsallis\xspace}
\newcommand{\GeVc}         {\ensuremath{\mathrm{Ge\kern-.1emV}/c}\xspace}
\newcommand{\MeVc}         {\mathrm{Me\kern-.1emV}/$c$\xspace}
\newcommand{\TeV}          {\mathrm{Te\kern-.1emV}\xspace}
\newcommand{\GeV}          {\mathrm{Ge\kern-.1emV}\xspace}
\newcommand{\MeV}          {\mathrm{Me\kern-.1emV}\xspace}
\newcommand{\GeVmass}      {\mathrm{Ge\kern-.2emV}/$c^2$\xspace}
\newcommand{\MeVmass}      {\mathrm{Me\kern-.2emV}/$c^2$\xspace}
\newcommand{\lumi}         {\ensuremath{\mathcal{L}_\mathrm{int}}\xspace}

\newcommand{\ITS}          {\rm{ITS}\xspace}
\newcommand{\TOF}          {\rm{TOF}\xspace}
\newcommand{\ZDC}          {\rm{ZDC}\xspace}
\newcommand{\ZDCs}         {\rm{ZDCs}\xspace}
\newcommand{\ZNA}          {\rm{ZNA}\xspace}
\newcommand{\ZNC}          {\rm{ZNC}\xspace}
\newcommand{\SPD}          {\rm{SPD}\xspace}
\newcommand{\SDD}          {\rm{SDD}\xspace}
\newcommand{\SSD}          {\rm{SSD}\xspace}
\newcommand{\TPC}          {\rm{TPC}\xspace}
\newcommand{\TRD}          {\rm{TRD}\xspace}
\newcommand{\VZERO}        {\rm{V0}\xspace}
\newcommand{\VZEROA}       {\rm{V0A}\xspace}
\newcommand{\VZEROC}       {\rm{V0C}\xspace}
\newcommand{\Vdecay} 	   {\ensuremath{V^{0}}\xspace}

\newcommand{\ee}           {\ensuremath{\mathrm{e^{+}e^{-}}}\xspace} 
\newcommand{\pip}          {\ensuremath{\pi^{+}}\xspace}
\newcommand{\pim}          {\ensuremath{\pi^{-}}\xspace}
\newcommand{\kap}          {\ensuremath{\rm{K}^{+}}\xspace}
\newcommand{\kam}          {\ensuremath{\rm{K}^{-}}\xspace}
\newcommand{\pbar}         {\ensuremath{\rm\overline{p}}\xspace}
\newcommand{\kzero}        {\ensuremath{{\rm K}^{0}_{\rm{S}}}\xspace}
\newcommand{\lmb}          {\ensuremath{\Lambda}\xspace}
\newcommand{\almb}         {\ensuremath{\overline{\Lambda}}\xspace}
\newcommand{\Om}           {\ensuremath{\Omega^-}\xspace}
\newcommand{\Mo}           {\ensuremath{\overline{\Omega}^+}\xspace}
\newcommand{\X}            {\ensuremath{\Xi^-}\xspace}
\newcommand{\Ix}           {\ensuremath{\overline{\Xi}^+}\xspace}
\newcommand{\Xis}          {\ensuremath{\Xi^{\pm}}\xspace}
\newcommand{\Oms}          {\ensuremath{\Omega^{\pm}}\xspace}
\newcommand{\degree}       {\ensuremath{^{\rm o}}\xspace}

\newcommand {\red}[1]    {\textcolor{red}{#1}}
\newcommand {\green}[1]  {\textcolor{green}{#1}}
\newcommand {\blue}[1]   {\textcolor{blue}{#1}}
\newcommand {\dgreen}[1]    {\textcolor{darkgreen}{#1}}
\newcommand {\orange}[1]   {\textcolor{orange}{#1}}

\newcommand{\deutsch}[1]{\foreignlanguage{german}{#1}}

\newcommand{\CZ}[1]{\blue{{\textbf{CZ:} #1}}}
\newcommand{\PA}[1]{\red{{\textbf{PA:} #1}}}
\newcommand{\JN}[1]{\red{{\textbf{JN:} #1}}}
\newcommand{\MF}[1]{\red{{\textbf{MF:} #1}}}
\newcommand{\todo}[1]{\red{{\textbf{TODO:} #1}}}
\newcommand{\todoartur}[1]{\todo{#1 (Artur)}}
\newcommand{\todobenedikt}[1]{\todo{#1 (Benedikt)}}
\newcommand{\todocristina}[1]{\todo{#1 (Cristina)}}
\newcommand{\todoluigi}[1]{\todo{#1 (Luigi)}}
\newcommand{\todoluuk}[1]{\todo{#1 (Luuk)}}
\newenvironment{oldconvert}
{\textbf{POTENTIALLY CONVERT/UPDATE/USE:}\itshape\color{gray}}

\newcommand{\TBC}       {\textcolor{red}{\footnotesize \textsc{TBC}}}

\DeclareRobustCommand{\unit}[2][]{%
        \begingroup%
                \def\0{#1}%
                \expandafter%
        \endgroup%
        \ifx\0\@empty%
                \ensuremath{\mathrm{#2}}%
        \else%
                \ensuremath{#1\,\mathrm{#2}}%
        \fi%
        }
\DeclareRobustCommand{\unitfrac}[3][]{%
        \begingroup%
                \def\0{#1}%
                \expandafter%
        \endgroup%
        \ifx\0\@empty%
                \raisebox{0.98ex}{\ensuremath{\mathrm{\scriptstyle#2}}}%
                \nobreak\hspace{-0.15em}\ensuremath{/}\nobreak\hspace{-0.12em}%
                \raisebox{-0.58ex}{\ensuremath{\mathrm{\scriptstyle#3}}}%
        \else
                \ensuremath{#1}\,%
                \raisebox{0.98ex}{\ensuremath{\mathrm{\scriptstyle#2}}}%
                \nobreak\hspace{-0.15em}\ensuremath{/}\nobreak\hspace{-0.12em}%
                \raisebox{-0.58ex}{\ensuremath{\mathrm{\scriptstyle#3}}}%
        \fi%
}

%
%
\newcommand{\ie}{i.\,e.\;}
\newcommand{\eg}{e.\,g.\;}

\newcommand{\run}[1]{\textsc{Run\,#1}}

%
%

\newcommand{\dNdeta}{\ensuremath{\mathrm{d}N_\mathrm{ch}/\mathrm{d}\eta}\xspace}

%
%
\newlength{\smallerpicsize}
\setlength{\smallerpicsize}{70mm}
\newlength{\smallpicsize}
\setlength{\smallpicsize}{90mm}
\newlength{\mediumpicsize}
\setlength{\mediumpicsize}{120mm}
\newlength{\largepicsize}
\setlength{\largepicsize}{150mm}

\newcommand{\PICX}[5]{
   \begin{figure}[!hbt]
      \begin{center}
         \vspace{3ex}
         \includegraphics[width=#3]{#1}
         \caption[#4]{\label{#2} #5}
      \end{center}
   \end{figure}
}

\newcommand{\PICH}[5]{
   \begin{figure}[H]
      \begin{center}
         \vspace{3ex}
         \includegraphics[width=#3]{#1}
         \caption[#4]{\label{#2} #5}
      \end{center}
   \end{figure}
}

%
%
%
\newcommand{\figs}{Figs.\xspace}
\newcommand{\Figs}{Figures\xspace}
\newcommand{\eqn}{equation\xspace}
\newcommand{\Eqn}{Equation\xspace}
\newcommand{\figref}[1]{Fig.~\ref{#1}}
\newcommand{\figsref}[2]{Figs.~\ref{#1}--~\ref{#2}}
\newcommand{\Figref}[1]{Figure~\ref{#1}}
\newcommand{\tabref}[1]{Tab.~\ref{#1}}
\newcommand{\Tabref}[1]{Table~\ref{#1}}
\newcommand{\appref}[1]{App.~\ref{#1}}
\newcommand{\Appref}[1]{Appendix~\ref{#1}}
\newcommand{\secs}{Secs.\xspace}
\newcommand{\Secs}{Sections\xspace}
\newcommand{\secref}[1]{Sec.~\ref{#1}}
\newcommand{\Secref}[1]{Section~\ref{#1}}
\newcommand{\chaps}{Chaps.\xspace}
\newcommand{\Chaps}{Chapters\xspace}
\newcommand{\chapref}[1]{Chap.~\ref{#1}}
\newcommand{\Chapref}[1]{Chapter~\ref{#1}}
\newcommand{\lstref}[1]{Listing~\ref{#1}}
\newcommand{\Lstref}[1]{Listing~\ref{#1}}
%
%
\newcommand{\otoprule}{\midrule[\heavyrulewidth]}
\topfigrule
%
\newcommand {\stat}     {({\it stat.})~}
\newcommand {\syst}     {({\it syst.})~}
 \newcommand {\mom}       {\ensuremath{p}}
\newcommand {\pT}        {\pt}
\newcommand {\meanpT}    {\ensuremath{\langle p_{\mathrm{T}} \kern-0.1em\rangle}\xspace}
\newcommand {\mean}[1]   {\ensuremath{\langle #1 \kern-0.1em\rangle}\xspace}
\newcommand {\sqrtsNN}   {\ensuremath{\sqrt{s_{\textsc{NN}}}}\xspace}
\newcommand {\sqrts}     {\ensuremath{\sqrt{s}}\xspace}
\newcommand {\vf}        {\ensuremath{v_{\mathrm{2}}}\xspace}
\newcommand {\et}        {\ensuremath{E_{\mathrm{t}}}\xspace}
\newcommand {\mT}        {\ensuremath{m_{\mathrm{T}}}\xspace}
\newcommand {\mTmZero}   {\ensuremath{m_{\mathrm{T}} - m_0}\xspace}
\newcommand {\minv}      {\mbox{$m_{\ee}$}}
\newcommand {\rap}       {\mbox{$y$}}
\newcommand {\absrap}    {\mbox{$\left | y \right | $}}
\newcommand {\rapXi}     {\mbox{$\left | y(\rmXi) \right | $}}
\newcommand {\abspseudorap} {\mbox{$\left | \eta \right | $}}
\newcommand {\pseudorap} {\mbox{$\eta$}}
\newcommand {\cTau}      {\ensuremath{c\tau}}
\newcommand {\sigee}     {$\sigma_E$/$E$}
\newcommand {\dNdpt}     {\ensuremath{\mathrm{d}N/\mathrm{d}\pT }}
\newcommand {\dNdptdy}   {\ensuremath{\mathrm{d^{2}}N/\mathrm{d}\pT\mathrm{d}y }}
\newcommand {\fracdNdptdy}   {\ensuremath{ \frac{\mathrm{d^{2}}N}{\mathrm{d}\pT\mathrm{d}y } }}
\newcommand {\dNdmtdy}   {\ensuremath{\mathrm{d^{2}}N/\mathrm{d}\mT\mathrm{d}y }}
\newcommand {\dN}        {\ensuremath{\mathrm{d}N }}
\newcommand {\dNsquared} {\ensuremath{\mathrm{d^{2}}N }}
\newcommand {\dpt}       {\ensuremath{\mathrm{d}\pT }}
\newcommand {\dy}        {\ensuremath{\mathrm{d}y}}
\newcommand {\dNdyBold}  {\ensuremath{\boldsymbol{\dN/\dy}}\xspace}
\newcommand {\dNchdy}    {\ensuremath{\mathrm{d}N_\mathrm{ch}/\mathrm{d}y }\xspace}
\newcommand {\dNchdeta}  {\ensuremath{\mathrm{d}N_\mathrm{ch}/\mathrm{d}\eta }\xspace}
\newcommand {\dNchdptdeta}  {\ensuremath{\mathrm{d}N_\mathrm{ch}/\mathrm{d}\pT\mathrm{d}\eta }\xspace}
\newcommand {\Raa}       {\ensuremath{R_\mathrm{AA}}}
\newcommand {\Nevt}      {\ensuremath{N_\mathrm{evt}}}
\newcommand {\NevtINEL}  {\ensuremath{N_\mathrm{evt}(\textsc{inel})}}
\newcommand {\NevtNSD}   {\ensuremath{N_\mathrm{evt}(\textsc{nsd})}}
\newcommand{\ttof}       {\ensuremath{t_\mathrm{TOF}}\xspace}
\newcommand {\ep}        {\mbox{$\mathrm {e\kern-0.05em p}$}\xspace}

\newcommand {\pplong}        {\mbox{\textit{proton--proton}}\xspace}
\newcommand {\ppBoldMath} {\mbox{$\mathrm { \mathbf p\kern-0.05em \mathbf p }$}\xspace}

\newcommand {\CuCu}      {\ensuremath{\mbox{Cu--Cu}}\xspace}
\renewcommand {\AA}      {\ensuremath{\mbox{A--A}}\xspace}

\newcommand {\Pbp}       {\ensuremath{\mbox{Pb--p}}\xspace}
\newcommand {\hPM}       {\ensuremath{h^{\pm}}\xspace}
\newcommand {\rphi}      {\ensuremath{(r,\phi)}\xspace}
\newcommand {\alphaS}    {\ensuremath{ \alpha_s}\xspace}
\newcommand {\MeanNpart} {\mbox{\ensuremath{< \kern-0.15em N_{part} \kern-0.15em >}}}

\newcommand {\sig}       {\ensuremath{S}\xspace}
\newcommand {\expsig}    {\ensuremath{\hat{S}}\xspace}
\newcommand {\prob}      {\ensuremath{P}\xspace}
\newcommand {\prior}     {\ensuremath{C}\xspace}
\newcommand {\prop}      {\ensuremath{F}\xspace}
\newcommand {\atrue}     {\ensuremath{\vec{A}_{\mathrm{true}}}\xspace}
\newcommand {\ameas}     {\ensuremath{\vec{A}_{\mathrm{meas}}}\xspace}
\newcommand {\detresp}   {\ensuremath{R}\xspace}

\newcommand {\pid}       {\ensuremath{\mathrm{\epsilon}_\mathrm{PID}}\xspace}
\newcommand {\nsigma}    {\ensuremath{\mathrm{n_{\sigma}}}\xspace}

\newcommand {\vzeroperc}        {\mbox{$\mathrm {V0M}_\mathrm{perc}$}\xspace}
\newcommand {\ntrkl}        {\mbox{$n_\mathrm{trkl}$}\xspace}

%
%
\newcommand {\mass}         {\mbox\mathrm{MeV$\kern-0.15em /\kern-0.12em c^2$}}
\newcommand {\tev}          {\ensuremath{\mathrm{TeV}}\xspace}
\newcommand {\gev}          {\ensuremath{\mathrm{GeV}}\xspace}
\newcommand {\mev}          {\ensuremath{\mathrm{MeV}}\xspace}
\newcommand {\kev}          {\ensuremath{\mathrm{keV}}\xspace}
\newcommand {\tevBoldMath}  {\ensuremath{\mathrm{\pmb{TeV}}}}
\newcommand {\gevBoldMath}  {\ensuremath{\mathrm{\pmb{GeV}}}}
\newcommand {\mmom}         {\ensuremath{\mathrm{MeV\kern-0.15em /\kern-0.12em c}}\xspace}
\newcommand {\gmom}         {\ensuremath{\mathrm{GeV\kern-0.15em /\kern-0.12em c}}\xspace}
\newcommand {\mmass}        {\ensuremath{\mathrm{MeV\kern-0.15em /\kern-0.12em c^2}}\xspace}
\newcommand {\gmass}        {\ensuremath{\mathrm{GeV\kern-0.15em /\kern-0.12em c^2}}\xspace}
\newcommand {\nb}           {\ensuremath{\mathrm{nb}}\xspace}
\newcommand {\musec}        {\ensuremath{\upmu \mathrm{s}}\xspace}
\newcommand {\nsec}         {\ensuremath{\mathrm{ns}}\xspace}
\newcommand {\psec}         {\ensuremath{\mathrm{ps}}\xspace}
\newcommand {\fmC}          {\ensuremath{\mathrm{fm}/c}\xspace}
\newcommand {\fm}           {\ensuremath{\mathrm{fm}}\xspace}
\newcommand {\mim}          {\ensuremath{\upmu\mathrm{m}}\xspace}
\newcommand {\cmq}          {\ensuremath{\mathrm{cm}^{2}}\xspace}
\newcommand {\mmq}          {\ensuremath{\mathrm{mm}^{2}}\xspace}
\newcommand {\dens}         {\ensuremath{\mathrm{g}/\mathrm{cm}^{3}}\xspace}
\newcommand {\lum}          {\ensuremath{\mathrm{cm}^{-2} \mathrm{s}^{-1}}\xspace}
\newcommand {\dg}           {\ensuremath{\kern+0.1em ^\circ}\xspace}
\newcommand{\mpp}           {\ensuremath{\mathrm{pp}}\xspace}
\newcommand{\rts}           {\ensuremath{\sqrt{s}}\xspace}
\newcommand{\gevc}          {\ensuremath{\mathrm{GeV}/c}\xspace}

\newcommand{\mevc}          {\ensuremath{\mathrm{MeV}/c}\xspace}
\newcommand{\mevcc}         {\ensuremath{\mathrm{MeV}/c^{2}}\xspace}
\newcommand{\gevcc}         {\ensuremath{\mathrm{GeV}/c^{2}}\xspace}

\newcommand{\kt}            {\ensuremath{k_\mathrm{T}}\xspace}

\newcommand{\nbinv}         {\ensuremath\mathrm{nb^{-1}}\xspace}
\newcommand {\ubinv}        {\ensuremath{\mu\rm b^{-1}}\xspace}
\newcommand{\mb}            {\ensuremath{\mathrm{mb}}\xspace}

\newcommand{\ctau}{\ensuremath{c\tau\xspace}}

%
%

\newcommand{\ePlusMinus}    {\ensuremath{\mathrm {e^{\pm}}}\xspace}
\newcommand{\muPlusMinus}   {\ensuremath{\upmu^{\pm}}\xspace}

\newcommand{\pion}          {\ensuremath{\uppi}\xspace}
\newcommand{\piZero}        {\ensuremath{\uppi^0}\xspace}
\newcommand{\piMinus}       {\ensuremath{\uppi^-}\xspace}
\newcommand{\piPlus}        {\ensuremath{\uppi^+}\xspace}
\newcommand{\piPlusMinus}   {\ensuremath{\uppi^\pm}\xspace}

\newcommand{\proton}        {\ensuremath{\mathrm{p}}\xspace}
\newcommand{\pOuPbar}       {\ensuremath{\mathrm {p^\pm}}\xspace}
\newcommand{\Bminus}        {\ensuremath{\mathrm{B^-}}\xspace}
\newcommand{\Bplus}         {\ensuremath{\mathrm{B^+}}\xspace}
\newcommand{\BZero}         {\ensuremath{\mathrm {B^0}}\xspace}
\newcommand{\BZerobar}      {\ensuremath{\mathrm {\overline{B}^0}}\xspace}

\newcommand{\rmLambdaZ}     {\ensuremath{\Lambda^0}\xspace}
\newcommand{\rmAlambdaZ}    {\ensuremath{\overline{\Lambda}^0}\xspace}
\newcommand{\rmLambda}      {\ensuremath{\Lambda}\xspace}
\newcommand{\rmAlambda}     {\ensuremath{\overline{\Lambda}}\xspace}
\newcommand{\rmLambdas}     {\ensuremath{\Lambda+\overline{\Lambda}}\xspace}

\newcommand{\Vzero}         {\ensuremath{\mathrm{V}^0}\xspace}
\newcommand{\Vzerob}        {\ensuremath{\mathrm{\pmb{V}}^0}\xspace}
\newcommand{\Kzero}         {\ensuremath{\mathrm{K}^0}\xspace}
\newcommand{\Kzs}           {\ensuremath{\mathrm{K_S}^0}\xspace}
\newcommand{\Ks}            {\Kzs}
\newcommand{\phimes}        {\ensuremath{\upphi}\xspace}
\newcommand{\Kminus}        {\ensuremath{\mathrm{K}^-}\xspace}
\newcommand{\Kplus}         {\ensuremath{\mathrm{K}^+}\xspace}
\newcommand{\Kstar}         {\ensuremath{\mathrm{K}^{*+}}\xspace}
\newcommand{\Kplusmin}      {\ensuremath{\mathrm{K}^\pm}\xspace}
\newcommand{\Jpsi}          {\ensuremath{\mathrm{J}/\uppsi}\xspace}
\newcommand{\PsiTwoS}       {\ensuremath{\uppsi(\mathrm{2S})}\xspace}
\newcommand{\DtoKpi}        {\ensuremath{\mathrm{D}^0 \to \mathrm{K}^-\uppi^+}\xspace}
\newcommand{\DtoKpipi}      {\ensuremath{\mathrm{D}^+\to \mathrm{K}^-\uppi^+\uppi^+}\xspace}
\newcommand{\DstartoDpi}    {\ensuremath{\mathrm{D}^{*+} \to \mathrm{D}^0 \uppi^+}\xspace}
\newcommand{\DstartoDpiToKpipi}    {\ensuremath{\mathrm{D}^{*+} \to \mathrm{D}^0 \uppi^+ \to \mathrm{K^-} \uppi^+ \uppi^+}\xspace}
\newcommand{\Dstophip}      {\ensuremath{\mathrm{D_{s}^+ \to \upphi \uppi^+ \to K^+K^-\uppi^+}}\xspace}
\newcommand{\DZero}         {\ensuremath{\mathrm{D}^0}\xspace}
\newcommand{\DZerobar}      {\ensuremath{\mathrm{\overline{D}^0}}\xspace}
\newcommand{\Dzero}         {\ensuremath{\mathrm{D^0}}\xspace}
\newcommand{\Dzerobar}      {\ensuremath{\overline{\mathrm{D}}^0}\xspace}
\newcommand{\Dstar}         {\ensuremath{\mathrm{D^{*+}}}\xspace}
\newcommand{\Dstarm}        {\ensuremath{\mathrm{D^{*-}}}\xspace}
\newcommand{\Dplus}         {\ensuremath{\mathrm{D^+}}\xspace}
\newcommand{\Dminus}        {\ensuremath{\mathrm{D^-}}\xspace}
\newcommand{\Ds}            {\ensuremath{\mathrm{D_{s}^+}}\xspace}
\newcommand{\Dsminus}       {\ensuremath{\mathrm{D_{s}^-}}\xspace}
\newcommand{\Lc}            {\ensuremath{\mathrm{\Lambda_{c}^+}}\xspace}
\newcommand{\Lcminus}       {\ensuremath\mathrm{{\overline{\Lambda}{}_c^-}}\xspace}
\newcommand{\Lcplus}        {\ensuremath\mathrm{{\Lambda_c^+}}\xspace}
\newcommand{\Xic}           {\ensuremath{\Xi_\mathrm{c}}\xspace}
\newcommand{\lambdab}       {\Lb}
\newcommand{\lambdac}       {\Lc}
\newcommand{\xicz}          {\ensuremath{\Xi_\mathrm{c}^0}\xspace}
\newcommand{\xiczp}         {\ensuremath{\Xi_\mathrm{c}^{0,+}}\xspace}
\newcommand{\xicp}          {\ensuremath{\Xi_\mathrm{c}^+}\xspace}
\newcommand{\xib}           {\ensuremath{\Xi_\mathrm{b}^0}\xspace}
\newcommand{\Lbzero}        {\Lb}
\newcommand{\LctopKpi}      {\ensuremath{\mathrm{\Lambda_{c}^{+}\to p K^-\uppi^+}}\xspace}
\newcommand{\LbtoLc}        {\ensuremath{\mathrm{\Lb \to \Lc + \mathrm{X}}}\xspace}
\newcommand{\LctopKzeros}   {\ensuremath{\mathrm{\Lc \to p \Kzs}}\xspace}
\newcommand{\KzStopippim}   {\ensuremath{\mathrm{\Kzs \to \uppi^+ \uppi^-}}\xspace}
\newcommand{\Lambdatoppim}  {\ensuremath{\mathrm{\Lambda \to p \uppi^{-}}}\xspace}

\newcommand{\decleng}       {\ensuremath{\mathrm{L_{xyz}}}\xspace}
\newcommand{\cosP}          {\ensuremath\mathrm{cos_{\Theta_{pointing}}}\xspace}

\newcommand{\ptLc}          {\ensuremath{p_\mathrm{T, \Lambda_c}}\xspace}
\newcommand{\ptpion}        {\ensuremath{p_\mathrm{T, \uppi}}\xspace}
\newcommand{\ptK}           {\ensuremath{p_\mathrm{T, K}}\xspace}
\newcommand{\ptproton}      {\ensuremath{p_\mathrm{T, \proton}}\xspace}

\newcommand{\da}            {\partial}
\newcommand{\de}            {\mathrm{d}}
\newcommand{\temp}[1]       {\textbf{\textcolor{red}{#1}}}
\newcommand{\slfrac}[2]     {\left.#1\right/#2}
\newcommand{\av}[1]         {\left\langle #1 \right\rangle}
\newcommand{\Gevc}          {\gevc}

\newcommand{\MeVcc}         {\mevcc}

\newcommand{\dm}            {\ensuremath{\Delta M}\xspace}

\newcommand{\mub}           {\ensuremath{\mathrm{\mu b}}\xspace}
\newcommand{\mum}           {\ensuremath{\mathrm{\mu m}}\xspace}
\newcommand{\DzerotoKpi}    {\DtoKpi}
\newcommand{\DplustoKpipi}  {\DtoKpipi}
\newcommand{\Dstophipi}     {\ensuremath{\mathrm{D_s^{+}\to \upphi\uppi^+}}\xspace}
\newcommand{\DstophipitoKKpi}{\Dstophip}
\newcommand{\phitoKK}       {\ensuremath{\mathrm{\upphi\to  K^+K^-}}\xspace}
\newcommand{\KKpi}          {\ensuremath{\mathrm{K^-\uppi^+\uppi^+}}\xspace}
\newcommand{\RAA}           {\ensuremath{R_\mathrm{AA}}\xspace}
\newcommand{\Ntrkl}         {\ensuremath{N_\mathrm{trkl}}\xspace}
\newcommand{\averNtrkl}     {\ensuremath{\langle N_\mathrm{trkl} \rangle}\xspace}
\newcommand{\Nch}           {\ensuremath{N_\mathrm{ch}}\xspace}
\newcommand{\averNch}       {\ensuremath{\langle N_\mathrm{ch} \rangle}\xspace}
\newcommand{\zVtx}          {\ensuremath{z_\mathrm{vtx}}\xspace}
\newcommand*\code[1]        {\texttt{#1}}
\newcommand{\fprompt}       {\ensuremath{f_\mathrm{prompt}}\xspace}
\newcommand{\fnonprompt}    {\ensuremath{f_\mathrm{non\text{-}prompt}}\xspace}
\newcommand{\AccEff}        {\ensuremath{\mathrm{Acc}\times\epsilon}\xspace}
\newcommand{\effP}[1]       {\ensuremath{\epsilon_#1^\mathrm{p}}\xspace}
\newcommand{\effNP}[1]      {\ensuremath{\epsilon_#1^\mathrm{np}}\xspace}
\newcommand{\rawY}[1]       {\ensuremath{Y_#1}\xspace}
\newcommand{\fP}            {\ensuremath{f_\mathrm{p}}\xspace}
\newcommand{\fNP}           {\ensuremath{f_\mathrm{np}}\xspace}

\newcommand{\HbtoDstar}     {\ensuremath{\mathrm{H_b\to\Dstar+X}}\xspace}
\newcommand{\fbtoHb}        {\ensuremath{f(\mathrm{b\to H_b})}\xspace}

\newcommand{\rzz}           {\ensuremath{\rho_{00}}\xspace}
\newcommand{\rzzo}          {\ensuremath{\rho_{00}^\mathrm{obs}}\xspace}
\newcommand{\rzzp}          {\ensuremath{\rho_{00}^\mathrm{prompt}}\xspace}
\newcommand{\rzznp}         {\ensuremath{\rho_{00}^\mathrm{non\text{-}prompt}}\xspace}
\newcommand{\rzznpevtg}     {\ensuremath{\rho_{00}^\mathrm{non\text{-}prompt,\evtgen}}\xspace}
\newcommand{\cost}          {\ensuremath{\cos{\vartheta^*}}\xspace}
\newcommand{\costsq}        {\ensuremath{\cos^2{\vartheta^*}}\xspace}
\newcommand{\costHel}       {\ensuremath{\cos{\vartheta^*}_\mathrm{helicity}}\xspace}
\newcommand{\costProd}      {\ensuremath{\cos{\vartheta^*}_\mathrm{production}}\xspace}
\newcommand{\ZZero}         {\ensuremath{\mathrm{Z}^0}\xspace}
\newcommand{\Lb}            {\ensuremath{\mathrm{{\Lambda_b^0}}}\xspace}
\newcommand{\pythia}        {\textsc{Pythia~8}\xspace}
\newcommand{\herwig}        {\textsc{Herwig~7}\xspace}
\newcommand{\geant}         {\textsc{Geant~4}\xspace}
\newcommand{\evtgen}        {\textsc{EvtGen}\xspace}
\newcommand{\lt}            {\ensuremath{\lambda_\vartheta}\xspace}

\newcommand{\DLL}           {\ensuremath{D^{\Dstar}_{1LL}(z)}\xspace}

\newcommand{\yi}            {\ensuremath{Y_{i}}\xspace}
\newcommand{\Np}            {\ensuremath{N_\mathrm{prompt}}\xspace}
\newcommand{\Nnp}           {\ensuremath{N_\mathrm{non\text{-}prompt}}\xspace}

%% file: intro.tex
\section{Introduction} 

The production of hadrons containing heavy charm and/or beauty quarks provides a powerful probe of quantum chromodynamics (QCD). Heavy quarks are predominantly produced in hard partonic-scattering processes, while their subsequent hadronisation into final-state particles is governed by non-perturbative dynamics. Although perturbative QCD calculations supplemented with fragmentation functions successfully describe heavy-flavour hadron production cross sections with continuously improving 
precision~\cite{ALICE:2023sgl,ALICE:2024xln}, understanding how the heavy-quark spin is transferred during hadronisation remains a much less constrained problem.
Polarisation and spin-alignment measurements therefore offer a sensitive and complementary probe of the fragmentation and decay dynamics of heavy-flavour hadrons~\cite{ALICE:2022byg,ALICE:2018crw,CMS:2024igk,LHCb:2023crj,LHCb:2025hul}. In relativistic heavy-ion collisions, these observables are of particular interest because non-central collisions generate a system with large orbital angular momentum, and strong electromagnetic fields being produced at early times~\cite{Becattini:2007sr,Kharzeev:2007jp,Skokov:2009qp,Deng:2012pc}. Such conditions make nucleus--nucleus collisions a unique environment in which spin-related observables can carry information on the initial geometry of the collision, the properties of the produced medium, and the mechanisms governing hadron formation. In recent years, vector-meson spin alignment has attracted considerable attention in this context~\cite{ALICE:2019aid, Chen:2026ton}. Besides its possible connection to the global angular momentum of the system, vector-meson spin alignment has also been discussed in relation to fluctuating colour fields~\cite{Kumar:2023xju}, hadronic rescattering~\cite{Li:2022vup}, spin-dependent hadronisation dynamics~\cite{Sheng:2023cql}, and holographic descriptions of strongly-coupled matter~\cite{Sheng:2024wzk}. The recent ALICE measurements of prompt \Dstar-meson spin alignment~\cite{ALICE:2025cdf} and \Jpsi polarisation~\cite{ALICE:2020iev,ALICE:2022sli} in Pb--Pb collisions have further strengthened this physics case by indicating that heavy-flavour vector mesons may exhibit spin-dependent effects in the presence of a strongly-interacting medium. These developments make precise measurements in smaller systems especially important for disentangling medium-induced effects from those already present in pp collisions and decay processes.

\begin{figure}[tb]
    \begin{center}
    \includegraphics[width = 0.8\textwidth]{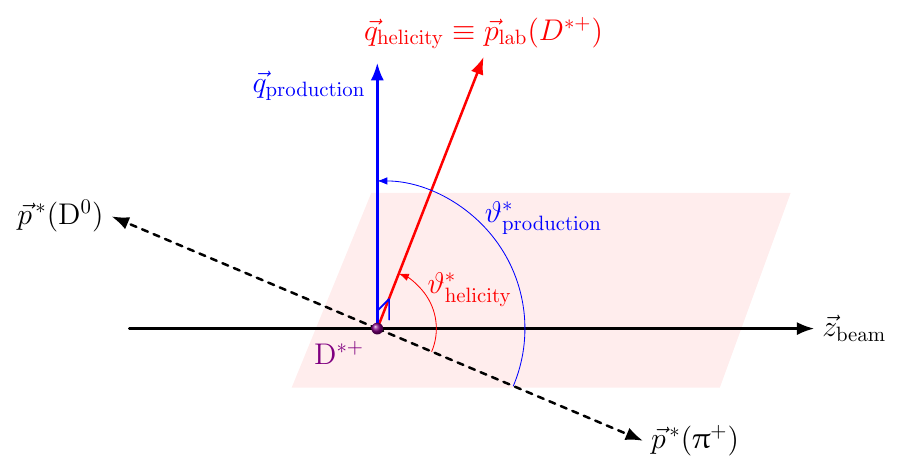}
    \end{center}
    \caption{Sketch of the \Dstar-meson decay and the reference frames considered for the spin alignment measurement, namely the helicity and production ones. The momentum vectors of the decay products \Dzero and $\uppi^+$ are considered in the rest frame of the parent \Dstar meson and the $\vartheta^*$ angles are between the momentum of the $\uppi^+$ and the two quantisation axes.}
    \label{fig-sketch}
\end{figure}

For a vector meson with spin $J=1$, spin alignment is quantified through the diagonal element \rzz of the spin-density matrix, defined with respect to a chosen quantisation axis~\cite{Fano:1957zz,Schilling:1969um}. This quantity gives the probability that the meson is produced in the spin projection state $m=0$ along that axis. In the absence of any preferred spin orientation, the three spin states $m=-1$, $0$, and $+1$ are populated equally, yielding $\rzz=1/3$. A deviation from this value indicates a non-uniform population of the spin substates, and hence a non-trivial tensor polarisation of the vector meson. Experimentally, $\rho_{00}$ is determined from the angular distribution of the decay products in the vector-meson rest frame. In particular, the distribution of the polar angle $\vartheta^*$, defined as the angle between the momentum of one of the decay products and the chosen quantisation axis, is given by
\begin{equation}
\frac{{\rm d}N}{{\rm d}\cos\vartheta^*} \propto (1-\rho_{00}) + (3\rho_{00}-1)\cos^2\vartheta^*.
\label{eq:rzz}
\end{equation} The value of $\rzz=1/3$ corresponds to no spin alignment, while values above or below $1/3$ correspond to an enhanced or reduced population of the $m=0$ state, respectively. The measured value of \rzz can depend on the choice of reference quantisation axis. In pp collisions, the helicity axis, defined by the vector-meson momentum in the laboratory frame, offers a natural choice for studying production and fragmentation effects~\cite{ALICE:2022byg}. The production axis, defined by the normal to the plane spanned by the beam direction and the vector-meson momentum, provides a complementary quantisation direction. While the helicity axis is used to define the spin alignment with respect to the meson momentum, the production axis is more directly sensitive to the geometry of the underlying scattering process. Comparing \rzz measured with respect to the two axes therefore helps assess the dynamical origin of the observed alignment and possible frame-dependent effects; moreover, in heavy-ion collisions the production plane becomes statistically correlated with the event plane in the presence of anisotropic flow, making production-axis measurements in pp collisions a natural baseline for event-plane-dependent studies~\cite{Faccioli:2010kd,ALICE:2022byg}. Figure~\ref{fig-sketch} shows a sketch of these two reference frames, and the $\vartheta^*$ angles with respect to the two quantisation axes.

Spin-alignment measurements of heavy-flavour vector mesons provide access to spin-dependent aspects of hadron formation that are not constrained by production yields and cross sections alone. 
The spin state of a vector meson can be determined experimentally from the angular distributions of its decay products, making spin alignment a sensitive probe of the interplay between production, hadronisation, and decay dynamics.
Among open-heavy-flavour mesons, the \Dstar meson is a particularly suitable probe~\cite{ALICE:2022byg,Belle:2019ewo,LHCb:2021sqa}. As a vector charm meson with experimentally accessible decay kinematics, it allows a precise determination of \rzz. Promptly produced \Dstar mesons probe the hadronisation of charm quarks into a spin-1 state, while non-prompt \Dstar mesons, originating from beauty-hadron decays, provide access to additional spin effects induced by weak decays. In the latter case, a deviation of \rzz from $1/3$ can arise naturally from helicity conservation and from the $V\!-\!A$ structure of the weak interaction in the $\mathrm{b}\to \mathrm{c}$ transition. The separation of prompt and non-prompt contributions is therefore essential for a meaningful interpretation of the measured spin alignment, since the two components probe distinct physical mechanisms~\cite{ALICE:2022byg}.

Measurements in pp collisions provide a direct laboratory for investigating spin-dependent hadronisation mechanisms in QCD, while also serving as an essential reference for heavy-ion studies.
Experimental studies in \ee and hadronic collisions have placed important constraints on the spin alignment of vector mesons, and theoretical approaches based on spin-dependent fragmentation functions have developed a quantitative framework for interpreting such effects~\cite{Donoghue:1978yb,Chen:2020pty}. 
In the fragmentation formalism, the spin alignment of a vector meson produced in a high-energy collision is governed by the tensor-polarised fragmentation function $D_{1LL}$~\cite{Chen:2020pty}, which encodes how a parton fragments into a spin-1 hadron with a non-uniform population of its spin substates. In this picture, the observable \rzz is directly related to the spin dependence of the fragmentation process itself. A particularly important consequence is that a value of \rzz other than $1/3$ does not require the fragmenting quark to be polarised: spin alignment can emerge even from the fragmentation of an unpolarised parton.
This makes measurements in pp collisions a direct probe of the spin-dependent fragmentation process, testing the spin-dependent fragmentation function that governs the population of vector-meson spin substates relative to the chosen quantisation axis, rather than serving only as a control measurement for heavy-ion studies.
Calculations in this framework, with $D_{1LL}$ constrained using \ee data, provide quantitative expectations for vector-meson spin alignment in hadronic collisions and predict that the effect can remain measurable in pp collisions~\cite{Chen:2020pty}. 

The first ALICE measurement of prompt and non-prompt \Dstar spin alignment in pp collisions at $\sqrt{s}=13~\TeV$ provided the first experimental test of these expectations~\cite{ALICE:2022byg}.
In that study, the diagonal spin-density-matrix element \rzz was measured at midrapidity with respect to the helicity axis. The prompt component was measured to be consistent with the unpolarised expectation, while the non-prompt component exhibited a clear deviation from $1/3$. The observed non-prompt \rzz behaviour was consistent with calculations based on PYTHIA 8~\cite{Sjostrand:2006za,Sjostrand:2014zea} coupled to the EvtGen decay package~\cite{Lange:2001uf}, where the weak-decay kinematics and helicity structure are modelled explicitly. These results demonstrated that prompt and non-prompt \Dstar mesons probe distinct physical mechanisms and that their experimental separation is crucial, particularly for future measurements in more complex collision systems.
They further established the role of event-generator calculations in disentangling the contributions from fragmentation and weak-decay dynamics to the observed spin alignment pattern~\cite{ALICE:2022byg,Sjostrand:2014zea,Lange:2001uf}.

In this paper, the measurement of the spin alignment of \Dstar mesons in pp collisions at $\sqrt{s}=13.6~\TeV$ obtained with ALICE is presented using LHC Run~3 data. 
Compared with the previous measurement at $\sqrt{s}=13~\TeV$ in the helicity frame, the present analysis benefits from the larger Run 3 data sample and the improved performance of the upgraded ALICE detector. This results in significantly better precision, an extended transverse-momentum reach, and the first extension of the measurement to the production-plane frame.
This allows testing a possible spin alignment of prompt \Dstar mesons with respect to a different quantisation axis, considering that \rzz was found to be compatible with $1/3$ in the helicity frame in the measurement reported in Ref.~\cite{ALICE:2022byg}.
It provides a more stringent test of spin-dependent heavy-quark fragmentation and weak-decay dynamics in pp collisions, strengthening the experimental baseline needed for the interpretation of heavy-ion results.
The results are compared with previous measurements~\cite{ALICE:2022byg}, with theoretical expectations based on spin-dependent fragmentation~\cite{Chen:2020pty}, and with Monte Carlo (MC) simulations using PYTHIA~8 coupled to the EvtGen decay package, which provide an essential benchmark for the description of prompt production and beauty-feed-down contributions. 

%% file: dataset.tex
\section{Experimental apparatus and data sample}

The ALICE experimental apparatus is composed of two main components: the central barrel covers the midrapidity region $|y|<0.9$, located inside a large solenoidal magnet, which provides a uniform magnetic field of $0.5$ T along the beam axis, and the muon spectrometer, which covers the forward rapidity range $-4<\eta<-2.5$~\cite{ALICE:2014sbx, Aamodt:2008zz}. During the second long shutdown of the LHC (LS2), the ALICE detector underwent a major upgrade programme in preparation for data taking in Runs 3 and 4~\cite{ALICE:2023udb}. This upgrade included, among other improvements, the installation of the new Inner Tracking System (ITS), the replacement of the Time Projection Chamber (TPC) readout chambers from multi-wire proportional chambers to gas electron multiplier (GEM) technology~\cite{Lippmann:2014lay,Adolfsson_2021}, and an upgrade of the Time-Of-Flight (TOF) readout electronics. The new ITS consists of seven layers of silicon detectors based on Monolithic Active Pixel Sensor (MAPS) technology~\cite{ALICE:2013nwm}, enabling a significantly improved determination of track parameters close to the interaction point and, consequently, the pointing resolution improved by a factor ranging between 2 and 5 depending on \pt and the particle direction, and a more precise reconstruction of primary and secondary vertices.
Starting with LHC Run 3, ALICE employs a continuous readout data-taking scheme, enabled in particular by the replacement of the TPC readout chambers with GEM-based technology, which allows ungated operation~\cite{ALICE:2023udb, Buncic:2015ari}. Under these conditions, 
the TPC provides up to 152 space points for the reconstruction of the charged-particle trajectory. In addition to tracking, it  also offers particle identification (PID) through the measurement of the specific ionisation energy loss $\mathrm{d}E/\mathrm{d}x$. The PID capabilities of the TPC are complemented by the TOF detector, which measures the flight time of charged particles from the interaction point~\cite{ALICE:2025tba}. The new Fast Interaction Trigger (FIT) detector~\cite{Trzaska:2017reu}, which includes the FV0, FT0, and FDD sub-detectors in the forward and backward regions, is used as a luminosity counter as well as a centrality estimator in heavy-ion collisions.

The upgrade permits the experiment to run at substantially higher interaction rates, reaching up to 2 MHz for pp collisions, while preserving excellent tracking capabilities and PID performance.
In this continuous readout scheme, multiple interactions are recorded within a single time frame~\cite{Buncic:2015ari}, making accurate association of collisions to bunch crossing (BC) essential. This association is achieved using the precise timing information provided by the FT0 detector with an intrinsic resolution of approximately $17~\mathrm{ps}$, which defines the minimum-bias (MB) trigger requirement through coincident signals in both forward FT0A ($3.5<\eta<4.9$) and backward FT0C ($-3.3 < \eta < -2.1$) detectors. In addition, the excellent vertex resolution provided by the upgraded ITS further enhances the separation of individual interactions and the correct association of tracks with their corresponding collision vertices. However, due to the finite time resolution, a fraction of reconstructed tracks may be compatible with more than one primary collision vertex. This ambiguity is addressed during the reconstruction by evaluating the compatibility of tracks with the different vertices based on their timing and spatial proximity. The excellent spatial resolution of the ITS is also crucial for reconstructing the decay vertices of heavy-flavour hadrons, which are typically displaced by a few hundred micrometres from the collision vertex where they are produced. The reconstruction of these displaced decays, together with the selection of characteristic decay-vertex topologies, provides a strong suppression of tracks incorrectly associated with the primary collision vertices.

The measurements are performed using the pp data sample collected at $\sqrt{s}=13.6 ~\TeV$ collected in 2022 and 2023, corresponding to an integrated luminosity of $\lumi \approx 6.49\pm0.15~\mathrm{pb}^{-1}$. The events were required to satisfy the MB condition described above, and the reconstructed primary vertices were required to be within $\pm10~\mathrm{cm}$ of the nominal interaction point along the beam direction. This trigger and vertex information was used to associate the collisions with the corresponding in-time BC. To avoid incomplete or inaccurately recorded collisions, events occurring close to the edges of the detector readout time frame were discarded. The fraction of events rejected due to this selection is about 20\%.

%% file: analysis.tex
\section{Data analysis}

\Dstar mesons and their charge conjugates are reconstructed at midrapidity ($|y|<0.8$) via the decay channel \DstartoDpiToKpipi, with branching ratio BR = $(2.67\pm0.03)\%$~\cite{ParticleDataGroup:2024}. 
The \Dzero-meson candidates are reconstructed by combining pairs of oppositely charged tracks reconstructed with the ITS and TPC detectors, assigning the kaon and pion mass hypotheses to the two tracks.
Each decay track is required to satisfy $|\eta|<0.8$ and $\pt >0.3~\GeVc$, and to have at least 70 (out of 152) associated space points in the TPC. In addition, a minimum of one hit in any of the three innermost layers of the ITS is required to ensure good pointing resolution. The \Dstar-meson candidates are reconstructed by combining the selected \Dzero-meson candidates with low \pt tracks, which are required to have $|\eta|< 0.8$, $\pt > 100~\MeVc$, and at least one hit in the ITS. The purity of the sample is improved by applying a selection on the invariant mass of the $\mathrm{K}\uppi$ pair forming the \Dzero candidate ($M(\Dzero)$), by requiring it to be compatible with the PDG value $M_\Dzero^\mathrm{PDG}$ within $125~\MeV/c^2$. Moreover, the \Dzero decay products are selected using the PID information provided by the TPC and TOF detectors. Pion and kaon candidates are selected by requiring the specific energy loss $\mathrm{d}E/\mathrm{d}x$ in the TPC detector and time of flight in the TOF detector to be compatible with the signal expected for the relevant particle hypothesis within three standard deviations.

\Dstar-meson candidates are selected using a multiclass classification algorithm based on Boosted Decision Trees (BDTs)~\cite{xgboost,hipe4ml}, in order to simultaneously suppress the large combinatorial background, as well as to separate the contributions of promt and non-prompt \Dstar mesons. Several topological and PID variables are used to train the BDT model, including: 1) the distance between the reconstructed \Dzero-meson decay vertex and the primary vertex, 2) the product of the signed transverse impact parameters of the \Dzero decay tracks with respect to the primary vertex, 3) the cosine of the pointing angle between the \Dzero-meson line of flight and its reconstructed momentum. Prompt and non-prompt \Dstar-meson signal samples used for the BDT training are obtained from MC simulations based on the PYTHIA 8.304 event generator using the beyond-leading colour approximation (BLC-CR) Mode 2 tune~\cite{Christiansen:2015yqa}, combined with the GEANT4 package for the transport of particles through the detector~\cite{AGOSTINELLI2003250}. The generated events are enriched with $\mathrm{c\bar{c}}$ and $\mathrm{b\bar{b}}$ pairs, and the decays of \Dstar mesons are forced into the channels of interest. 
The background samples are obtained from data in the region of the invariant-mass difference distribution, where $\Delta M=M_\Dstar-M_\Dzero$, the selected sideband region is $150<\Delta M~<170~\mevcc$, where no \Dstar-meson signal is expected. 
Independent BDTs are trained for the different \pt intervals of the analysis. 
The trained BDTs are subsequently applied to the data sample, producing three scores for each candidate. These scores quantify the likelihood of the candidate being a non-prompt \Dstar meson, prompt \Dstar meson or part of the combinatorial background, respectively. 
A selection on the BDT output score, related to the probability of being a background candidate, is applied to improve the signal-to-background ratio. The optimal working point is determined by scanning different cut values and selecting the one that maximises the signal significance, evaluated as $S/\sqrt{S+B}$, where $S$ and $B$ denote the numbers of signal and background candidates, respectively. 
In addition, a selection based on the BDT score, which is trained to discriminate between prompt and non-prompt candidates, is applied to define prompt(non-prompt)-enhanced samples.

\begin{figure}[tb]
    \begin{center}
    \includegraphics[width = 1\textwidth]{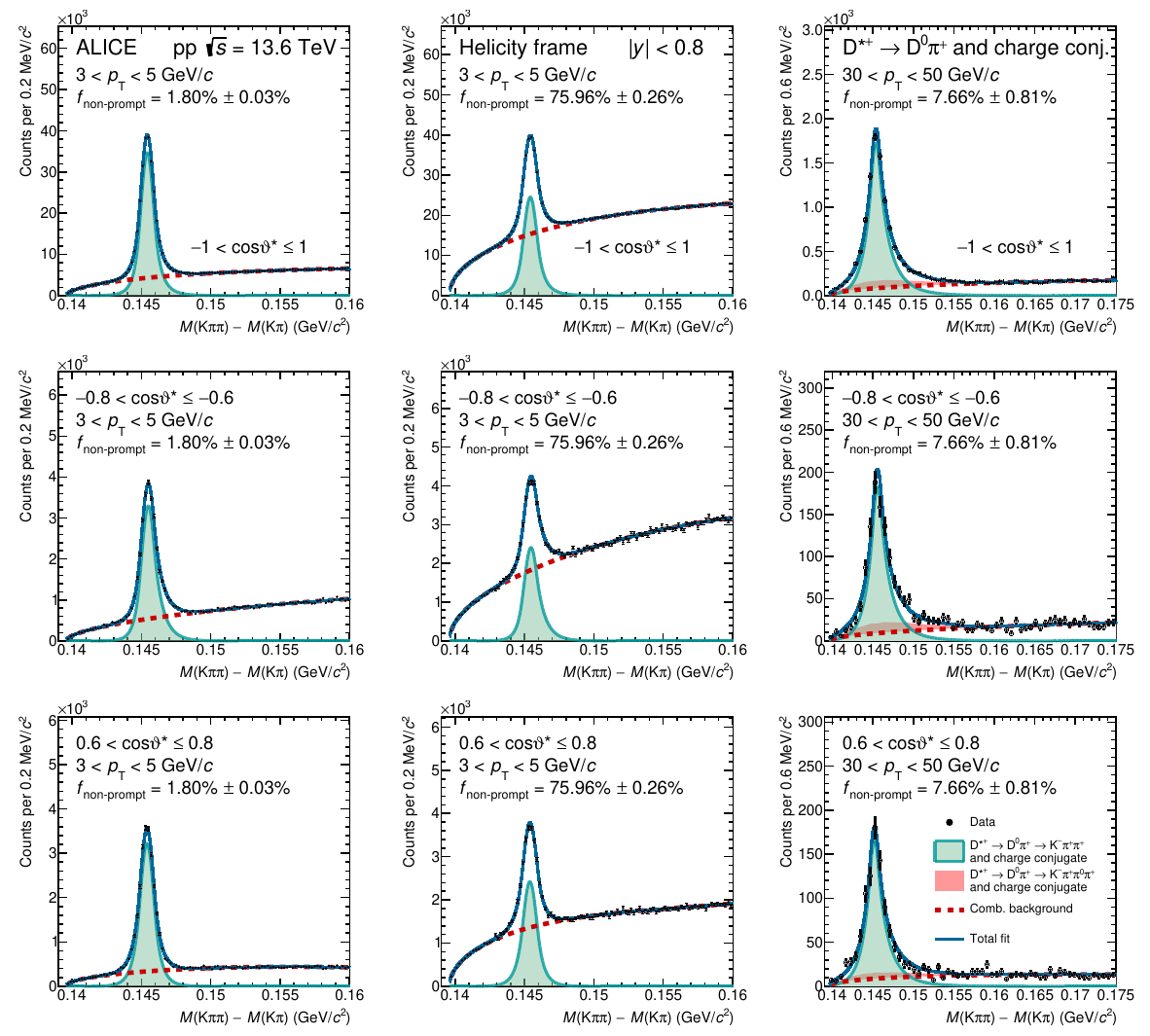}
    \end{center}
    \caption{Invariant-mass distributions $\Delta M$ of \Dstar-meson candidates obtained from prompt (left) and non-prompt (middle) enhanced sample in $3<\pt<5~\GeVc$ and for the prompt enhanced sample in $30<\pt<50~\GeVc$ (right). Distributions are shown for $\cost$ integrated (top), $-0.8<\cost<-0.6$ (middle) and $0.6<\cost<0.8$ (bottom) at midrapidity ($|y|<0.8$) in pp collisions at $\sqrt{s}=13.6$~Te\kern-.1emV\xspace.}
    \label{fig-massfit}
\end{figure}

The analysis is performed in seven \pt intervals within the range $3 < \pt < 100~\GeVc$: [3–5], [5–7], [7–10], [10–20], [20–30], [30-50] and [50–100] $\GeVc$. In each \pt interval, the raw yield of \Dstar mesons is extracted by fitting the $\Delta M$ distributions in ten \cost intervals covering the full range from -1 to 1 with equal width, where $\vartheta^*$ is the angle between the momentum vector of either the \Dzero meson or the pion track in the rest frame of the \Dstar meson and the quantisation axis. 
The $\Delta M$ distributions are fitted using a combination of a double-sided Crystal Ball function~\cite{Eschle:2019jmu} to describe the signal peak 
and a function to model the combinatorial background,
\begin{equation}
f(\Delta M)=N(\Delta M-M_\uppi)^{p_0}e^{p_1(\Delta M-M_\uppi)+p_2(\Delta M-M_\uppi)^2+p_3(\Delta M-M_\uppi)^3},
 \label{eq:background}
\end{equation}
where $N$, $p_0$, $p_1$, $p_2$ and $p_3$ are free parameters and $M_\uppi$ is the mass of the charged pion. Moreover, for $\pt>30~\GeVc$, an additional background source originating from \Dstar mesons decaying via the $\Dstar\rightarrow\Dzero\uppi^+$ channel, followed by $\Dzero\rightarrow \mathrm{K}^-\uppi^+\uppi^0$, is included in the fit. At high \pt, the \Dzero signal peak becomes significantly broader due to the worsening momentum resolution, leading to an increased overlap between different decay contributions. As a result, this contribution can not be completely removed by the $M(\Dzero)$ selection. Its $\Delta M$ distributions and the relative abundance with respect to \Dstar-meson signals are obtained from MC simulations. To further improve the fit stability at high \pt, the signal parameters are fixed to those extracted from the \cost-integrated $\Delta M$ distributions for $\pt>20~\GeVc$. 
Fig.~\ref{fig-massfit} shows representative $\Delta M$ distributions obtained in the helicity frame, the similar $\Delta M$ distributions are also obtained for the production frame. The left column corresponds to the sample with the largest prompt contribution (prompt enhanced) in $3<\pt<5~\GeVc$, the middle column shows the non-prompt enhanced sample in the same \pt interval, and the right column depicts the prompt enhanced sample in the interval $30<\pt<50~\GeVc$. The first row is integrated over the full angular range, $-1<\cost<1$. The second row corresponds to $-0.8<\cost<-0.6$, and the third row to $0.6<\cost<0.8$.

\begin{figure}[tb]
    \begin{center}
    \includegraphics[width = 1\textwidth]{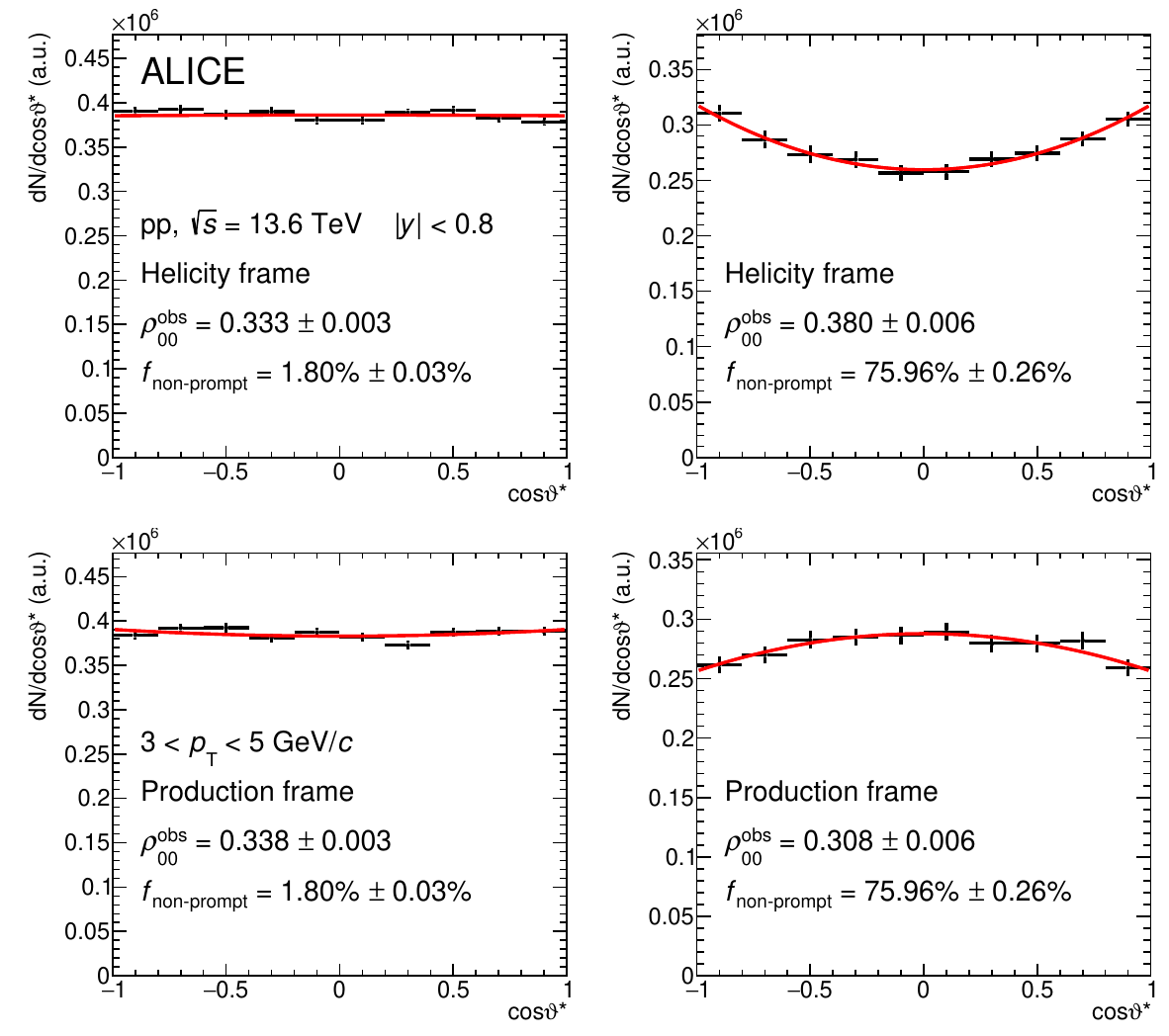}
    \end{center}
    \caption{Acceptance and efficiency corrected angular distribution of the decay products in the rest frame of the \Dstar mesons with respect to the helicity (upper panels) and production (bottom panels) frame. The left panels show the angular distribution for the charm-enhanced \Dstar sample, and the right panel shows the same but for the beauty-enhanced \Dstar sample. The distributions are obtained in $3<\pt<5~\GeVc$ and are fitted with Eq.~\ref{eq:rzz} to extract the value of spin density matrix element $\rzz^\mathrm{obs}$.}
    \label{fig-dist-yield}
\end{figure}

The \Dstar-meson raw yields are corrected for the product of the geometrical acceptance and the reconstruction and selection efficiency (\AccEff), in each \cost and \pt interval. The corresponding \AccEff is evaluated using an MC simulation, similar to the one used for BDT trainings. The generated MC distributions are reweighted to reproduce the predicted \pt distributions based on the FONLL prediction~\cite{Cacciari:1998it,Cacciari:2001td}. The resulting variation of the \AccEff values with respect to those obtained without reweighting is found to be negligible. 
Finally, the observed spin-alignment parameter $\rzz^\mathrm{obs}$ is extracted in each \pt interval with respect to the production and helicity axes separately, with the functional form given in Eq.~\ref{eq:rzz}.
The angular distributions corrected for efficiency and acceptance for the selected \pt interval with both helicity (upper panels) and production (bottom panels) axis are shown in Fig.~\ref{fig-dist-yield}. The left panels present the distributions for the prompt-enhanced \Dstar samples, and the right panels present the same for the non-prompt-enhanced \Dstar samples. The angular distributions are fitted with the functional form given in Eq.~\ref{eq:rzz}.

The raw yields contain a mixture of prompt and non-prompt \Dstar mesons. Therefore, the extracted \rzzo is expected to follow a linear combination of prompt (\rzzp) and non-prompt (\rzznp) contributions, which can be expressed as
\begin{equation}
   \rzz^\mathrm{obs} = \fprompt\times \rzzp+\fnonprompt\times \rzznp,
 \label{eq:extrap}
\end{equation}

where $f_\mathrm{(non\text{-})prompt}$ is the fraction of (non-)prompt \Dstar mesons~\cite{ALICE:2022byg, CMS:2020qul}. This fraction can be estimated via a data-driven approach~\cite{ALICE:2021mgk}. This approach is based on the sampling of the raw yield \yi at different values of the BDT score corresponding to the probability for a candidate to be a non-prompt \Dstar meson. \yi can be related to the corrected yields of the prompt (\Np) and non-prompt (\Nnp) \Dstar mesons by the factors \AccEff by
\begin{equation}
   (\AccEff)^{\rm prompt}_{i} \times \Np +(\AccEff)^{\rm non\text{-}prompt}_{i}\times \Nnp - \yi = \delta_{i},
 \label{eq:acceff}
\end{equation}

where $\delta _{i} $ represents the residual term originating from the uncertainties on $Y_{i}$, $(\AccEff)^{\rm prompt}_{i}$, and $(\AccEff)^{\rm non\text{-}prompt}_{i}$. The system of equations can be solved via a $\chi^2$-minimisation procedure, together with their covariance matrix.
For a given set $j$ of BDT selections, the fraction of (non-)prompt \Dstar mesons, $f_{\rm (non\text{-})prompt}$ is computed as
\begin{equation}
   f^{\rm (non\text{-})prompt}_{j}= \frac{(\AccEff)^{\rm (non\text{-})prompt}_{j}\times N_{(non\text{-})prompt} }{(\AccEff)^{\rm prompt}_{j}\times \Np+(\AccEff)^{\rm non\text{-}prompt}_{j}\times \Nnp}.
 \label{eq:frac}
\end{equation}

The uncertainty in \fnonprompt is determined by error propagating using the covariance matrix of \Np and \Nnp obtained from the minimisation of $\chi^2$.

\begin{figure}[tb]
    \begin{center}
    \includegraphics[width = 1\textwidth]{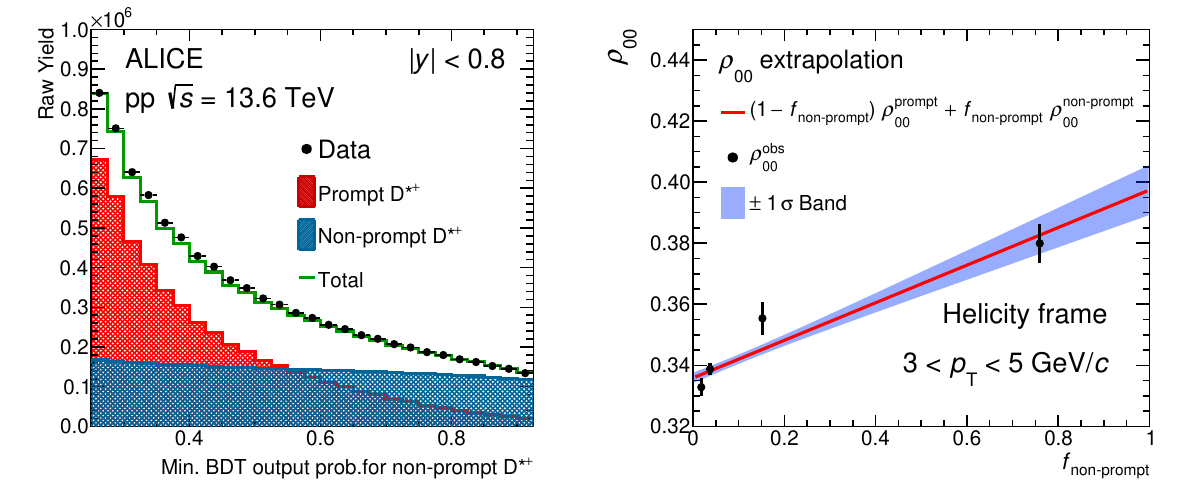}
    \end{center}
    \caption{Left: raw yield as a function of BDT selections and extracted contributions from prompt and non-prompt \Dstar mesons with $3<\pt<5~\GeVc$. Right: the spin density matrix element (\rzzo) with respect to the helicity axis as a function of $f_{\rm non\text{-}prompt}$ in $3<\pt<5~\GeV/c$ for charm- and beauty-enhanced samples. Only statistical uncertainties are shown. The red line represents the linear fit to the data with the blue band indicating its $1\sigma$ confidence interval.}
    \label{fig-cutvar}
\end{figure}

 The left panel of Fig.~\ref{fig-cutvar} presents an example of the raw yield as a function of the BDT selections for \Dstar mesons with $3<\pt<5~\GeVc$. The leftmost point corresponds to the loosest BDT requirement on the candidate probability of being a non-prompt \Dstar meson, while the rightmost point corresponds to the tightest selection, preferentially enhancing the non-prompt component. The prompt and non-prompt contributions are shown as the red and blue filled histograms, respectively, with their sum represented by the green histogram.

Finally, for $\pt<30~\GeV/c$, the distribution of the BDT score associated with the probability of being a non-prompt \Dstar meson is divided into four independent intervals. This defines four independent sub-samples with different non-prompt (prompt) contributions. The \rzzo is then extracted using these four non-overlapping BDT-score intervals and a linear fit based on Eq.~\ref{eq:extrap}. The \rzzp and \rzznp parameters are obtained by evaluating the fit at $\fnonprompt = 0$ and $\fnonprompt = 1$, respectively. The right panel of Fig.~\ref{fig-cutvar} shows an example of the linear fit of \rzzo with respect to the helicity axis as a function of \fnonprompt in the range $3<\pt<5~\GeVc$. The coloured band represents the $1\sigma$ confidence interval obtained from the linear fit. This procedure is not applied for $\pt>30~\GeV/c$ due to the limited size of the data sample. In this case, the \rzzp is obtained from Eq.~\ref{eq:extrap} by fixing the \rzznp value to the one obtained from PYTHIA 8 simulations coupled with the EvtGen decay package. Since the fraction of prompt \Dstar mesons in the measured sample is about 90\%, the subtraction of the residual non-prompt contamination results in a correction of the measured \rzz parameter smaller than $0.01$.

%% file: systematics.tex
\section{Systematic uncertainties}
The systematic uncertainties of the prompt and non-prompt \Dstar-meson \rzz are estimated for each \pt interval and for the helicity and production reference frames separately. The main sources are (i) the signal extraction, (ii) the track reconstruction and selection efficiencies, (iii) the (non-)prompt fraction estimation, and (iv) the BDT selection efficiency. All sources of systematic uncertainty are considered to be uncorrelated and the total systematic uncertainty is obtained by summing the different contributions in quadrature. While the different sources of systematic uncertainty are largely independent, a potential correlation may exist between the uncertainties associated with the BDT selection efficiency and the (non-)prompt fraction determination, since the latter is estimated using the efficiencies corresponding to different BDT output score selections. However, the systematic uncertainty related to the (non-)prompt fraction determination is negligible compared with the other contributions. Consequently, any residual correlation with the other sources would have a negligible impact on the total systematic uncertainty.
The values of total systematic uncertainties range between 2.5\% and 6.2\%, depending on \pt. The estimated values for representative \pt intervals are summarised in Table~\ref{tab:sys}.

\begin{table}[!b]
    \centering
    \caption{Summary of systematic uncertainties assigned for the measurement of the \Dstar-meson spin alignment in pp collisions at $\s=13.6~\TeV$ for different \pt intervals.}
    \resizebox{\textwidth}{!}{
    \begin{tabular}{c|c|c|c|c|c|c}
        \hline
        \multicolumn{1}{c}{}&\multicolumn{3}{c}{Helicity axis} & \multicolumn{3}{c}{Production axis}\\
        \cline{2-7}
        \multicolumn{1}{c}{}&\multicolumn{2}{c|}{Prompt} & Non-prompt &\multicolumn{2}{c|}{Prompt} & Non-prompt \\
        \cline{2-4}\cline{5-7}
         \multicolumn{1}{c}{$\pt~(\GeV/c)$} & $3\text{--}5$ & $50\text{--}100$ & $3\text{--}5$ & $3\text{--}5$ & $50\text{--}100$ & $3\text{--}5$ \\
        \hline
        Signal yield (\%)           & \multicolumn{1}{c}{1}   & \multicolumn{1}{c}{2}   & \multicolumn{1}{c|}{1}     & \multicolumn{1}{c}{1}     & \multicolumn{1}{c}{3}     & \multicolumn{1}{c}{2}   \\
     
        Track efficiency(\%)        & \multicolumn{1}{c}{1}     & \multicolumn{1}{c}{5}     & \multicolumn{1}{c|}{3}     &  \multicolumn{1}{c}{1}    & \multicolumn{1}{c}{5}     & \multicolumn{1}{c}{1.5}   \\

       (Non-)prompt fraction (\%)  & \multicolumn{1}{c}{0.1}   & \multicolumn{1}{c}{0.5}   & \multicolumn{1}{c|}{0.2}   & \multicolumn{1}{c}{0.1}   & \multicolumn{1}{c}{0.2}   & \multicolumn{1}{c}{0.1}   \\
         
        BDT efficiency (\%)         & \multicolumn{1}{c}{2}     & \multicolumn{1}{c}{2}     & \multicolumn{1}{c|}{2}     & \multicolumn{1}{c}{2}     & \multicolumn{1}{c}{2}     & \multicolumn{1}{c}{2}   \\
        \hline
        Total (\%)                  & \multicolumn{1}{c}{2.5}     & \multicolumn{1}{c}{5.8}  & \multicolumn{1}{c|}{3.7}   & \multicolumn{1}{c}{2.5}   & \multicolumn{1}{c}{6.2}   & \multicolumn{1}{c}{3.2}  \\
        \hline
    \end{tabular}
    }
    
    \label{tab:sys}
\end{table}

The systematic uncertainty of the signal extraction is evaluated by repeating the invariant-mass fit procedure under different conditions: varying the lower and upper limits of the fit range, using an alternative functional form for the background description, and leaving the signal peak width free in the fits. In addition, the bin-counting method is also used to estimate a systematic uncertainty. In this procedure, the signal yield is determined by integrating the invariant-mass distribution within a selected window after background subtraction. Different mass ranges are considered for the bin-counting and the resulting variations in \rzz are propagated as a systematic uncertainty. 
For each source, the systematic uncertainty is taken as the RMS of the distribution of the differences between the variation results and the default result.
 
The values of systematic uncertainty range from 1\% to 2\% with respect to the helicity axis, and 1\% to 3\% with respect to the production axis, depending on \pt.

The track selection and reconstruction efficiency systematic uncertainty is estimated by varying the track-quality selection criteria and testing the variation of the final \rzz parameters. The variations applied specifically concern the minimum number of ITS and TPC clusters for decay tracks and their $\mathrm{\eta}$ acceptance. These variations account for possible mismodelling of the detector response, including effects related to the alignment and resolution, and their impact on the angular distributions of the decay products. The systematic uncertainty ranges between 1\% to 5\% with respect to both the helicity and production axes, depending on \pt.

The uncertainty on the determination of the prompt and non-prompt fractions is assessed by varying the selection criteria in the minimisation procedure used to separate the two components, as well as by modifying the invariant-mass fit configurations employed in the extraction of the corresponding yields. With respect to the helicity axis, the systematic uncertainty varies from 0.1\% to 0.5\%, while for the production axis it ranges from 0.1\% to 0.2\%, depending on \pt.

The systematic uncertainty related to the BDT selection efficiency originates from imperfections in the simulation of the decay kinematics and from residual differences between data and MC for the variables used in the BDT, particularly those related to the impact parameter and momentum resolution. It is quantified by repeating the analysis with different BDT working points and comparing the extracted \rzz values for both prompt and non-prompt \Dstar mesons. The values of systematic uncertainties are uniformly 2\% across all \pt and with respect to both the helicity and production axes.

As an additional consistency check, the analysis was repeated using a random quantisation axis, defined by a unit vector with random orientation in three-dimensional space. This procedure removes any physical correlation between the chosen axis and the momentum direction of the decay products. The resulting \rzz values for both prompt and non-prompt \Dstar mesons are found to be consistent with the unpolarised expectation of $1/3$, validating the robustness of the measurements.

%% file: results.tex
\section{Results}

\begin{figure}[!tb]
    \begin{center}
    \includegraphics[width = 1\textwidth]{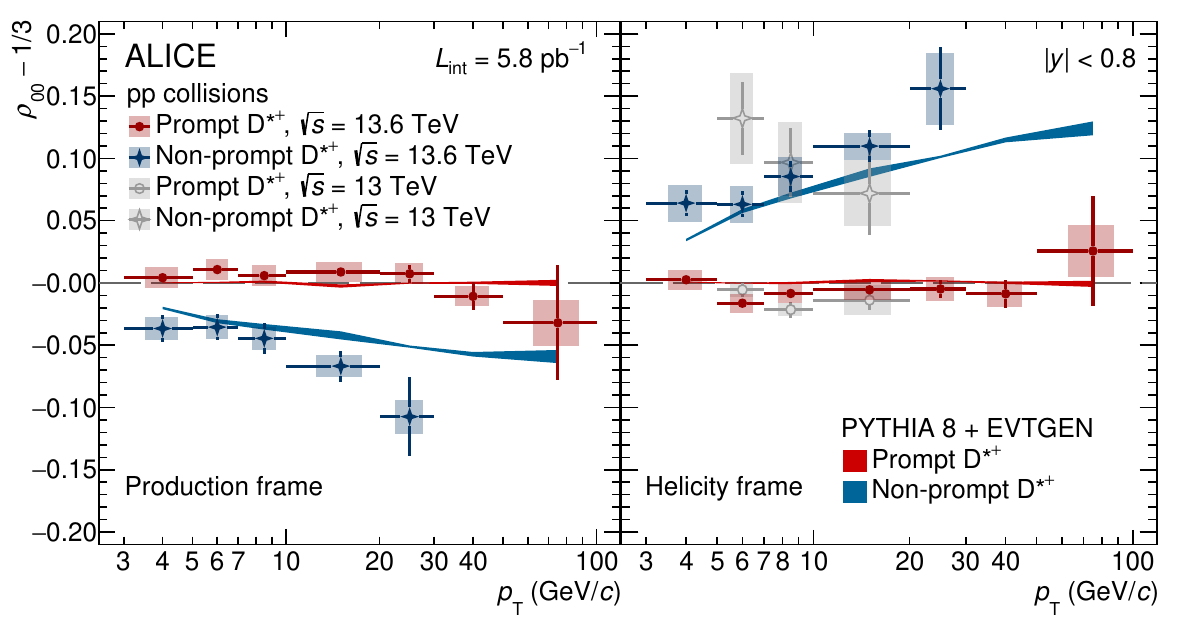}
    \end{center}
    \caption{The spin density matrix element (\rzz) for prompt and non-prompt \Dstar mesons as a function of \pt in the production (left panel) and helicity (right panel) frames in pp collisions at $\s=13.6~\TeV$. The measurements in the helicity frame are compared with previous results at $\s=13~\TeV$~\cite{ALICE:2022byg}. All the data points are compared with predictions from the PYTHIA 8 + EvtGen MC generators~\cite{Sjostrand:2006za,Sjostrand:2014zea, Lange:2001uf}.}
    \label{fig-results}
\end{figure}

Figure~\ref{fig-results} presents the measured values of the spin-density matrix element \rzz for prompt and non-prompt \Dstar mesons at midrapidity ($|y|<0.8$) as a function of \pt in pp collisions at $\s=13.6~\TeV$. The results are reported with respect to both the production (left panel) and the helicity (right panel) axes. The data points are compared to the measurement with respect to the helicity axis performed in pp collisions at $\s=13~\TeV$~\cite{ALICE:2022byg}.

Within uncertainties, the prompt \Dstar-meson \rzz is consistent with the unpolarised expectation of $\rzz=1/3$ over the entire measured \pt range up to $100~\Gevc$ in both the helicity and production frames. These results extend the previous measurement performed in pp collisions at $\s=13~\TeV$ to substantially higher \pt with improved statistical precision and confirm the absence of spin alignment for promptly produced \Dstar mesons in pp collisions. The observations are consistent with previous measurements in \ee collisions and indicate that possible charm-quark polarisation effects are either negligible at production or largely diluted during the hadronisation process~\cite{CLEO:1991bnf, TPCTwo-Gamma:1991mvk, HRS:1987vlf}. In contrast, the \rzz values measured for non-prompt \Dstar mesons with respect to the helicity (production) axis are systematically larger (smaller) than $1/3$ across the studied \pt range. This behaviour is expected as a consequence of helicity conservation in the weak decays of beauty hadrons to vector charm mesons, which are dominated by scalar B-meson contributions. The results obtained in this work are consistent with those obtained at $\s=13~\TeV$, while benefitting from an extended kinematic range, and providing a coherent picture in the two different reference frames.
The measurements are compared to MC simulations based on PYTHIA 8 coupled with the EvtGen decay package~\cite{Sjostrand:2006za,Sjostrand:2014zea, Lange:2001uf}. In these simulations, EvtGen explicitly accounts for helicity conservation and the $V\!-\!A$ structure of the weak interaction in beauty-hadron decays. The predicted \rzz values are found to be in good agreement with the experimental results for both prompt and non-prompt \Dstar mesons in all analysed reference frames. When EvtGen is not used, the simulated \rzz values are compatible with $1/3$ for both components, highlighting the essential role of the decay modelling in reproducing the observed spin alignment for non-prompt \Dstar mesons.

\begin{figure}[!tb]
    \begin{center}
    \includegraphics[width = 0.7\textwidth]{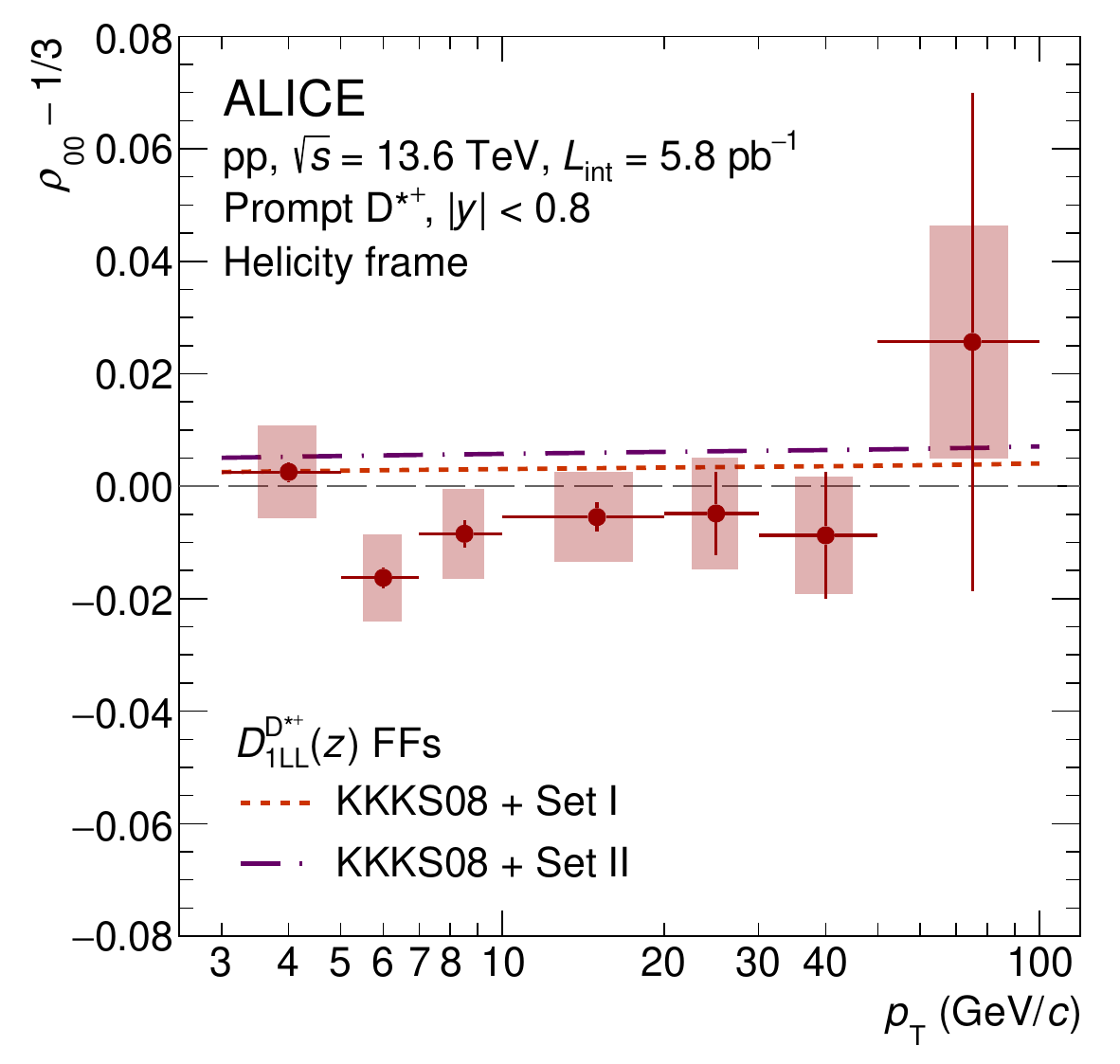}
    \end{center}
    \caption{The spin density matrix element \rzz for prompt \Dstar mesons as a function of \pt with respect to the helicity axis in pp collisions at $\s=13.6~\TeV$. The measurements are compared with the theoretical calculations with spin-dependent fragmentation functions \DLL (FFs) tuned to measurements from \ee collisions~\cite{Chen:2020pty}. No uncertainties are considered in the model predictions.}
    \label{fig-withFFs}
\end{figure}

The measurement of the prompt \Dstar-meson \rzz with respect to the helicity axis is further compared with theoretical predictions based on spin-dependent fragmentation functions (FFs) tuned to measurements in \ee collisions~\cite{Chen:2020pty}, as shown in Fig.~\ref{fig-withFFs}. This model is based on perturbative QCD (pQCD) calculations carried out within the collinear factorisation scheme. Within this approach, the spin-dependent differential cross section is computed as a convolution of three terms, (i) the parton distribution functions (PDFs) of the incoming protons, (ii) the partonic cross section expressed as a perturbative series expansion in the strong coupling constant, and (iii) the spin-dependent FFs, \DLL. The partonic cross section is evaluated up to the perturbative order that incorporates the first-order perturbative QCD evolution of the FFs. The spin-dependent fragmentation functions \DLL are based on the unpolarised FFs KKKS08~\cite{Kneesch:2007ey} and the spin-dependent term is constrained to measurements performed in \ee collisions. In particular, two sets of measurements were considered. The first one (Set I) includes measurements from the CLEO II~\cite{CLEO:1998lwi}, HRS~\cite{HRS:1987vlf}, TPC~\cite{TPCTwo-Gamma:1991mvk}, SLD~\cite{SLD:1997grw}, and OPAL~\cite{OPAL:1997nwj} experiments, while in the second one (Set II), the results of the OPAL experiment were excluded, since the measurement of non-prompt \Dstar mesons is not consistent with those reported by the other experiments. In both configurations, the predicted \rzz of prompt \Dstar mesons is slightly higher than $1/3$, without showing a sizeable dependence on the \Dstar-meson \pt. The experimental uncertainties of the \ee measurements are not considered in the theoretical predictions. The central values of the measured \rzz are instead slightly lower than $1/3$, but still compatible with both predictions based on the spin-dependent FFs within about one standard deviation.

\begin{figure}[!tb]
    \begin{center}
    \includegraphics[width = 0.7\textwidth]{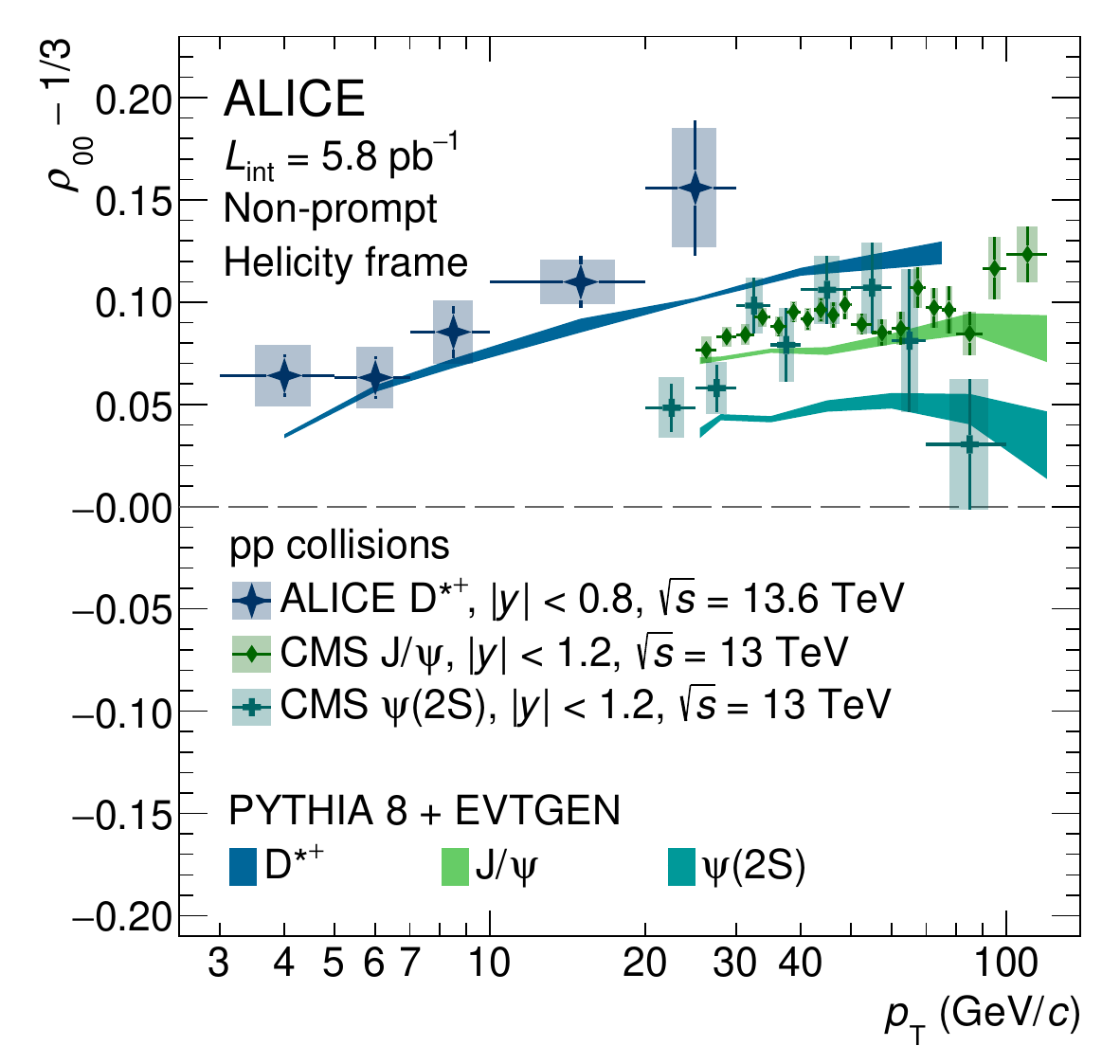}
    \end{center}
    \caption{The spin density matrix element \rzz of non-prompt \Dstar mesons with respect to the helicity axis as a function of \pt in the rapidity interval $|y| < 0.8$ in pp collisions at $\s=13.6~\TeV$, compared to that of non-prompt \Jpsi and \PsiTwoS mesons in the rapidity interval $|y|<1.2$ measured by the CMS Collaboration in pp collisions at $\s=13~\TeV$~\cite{CMS:2024igk}. All the data points are also compared to predictions from the PYTHIA 8 + EvtGen MC generators~\cite{Sjostrand:2006za,Sjostrand:2014zea, Lange:2001uf}.}
    \label{fig-withJpsi}
\end{figure}

Models describing prompt charmonium production incorporate polarisation effects in the pre-resonance $\mathrm{Q\overline{Q}}$ states, which are the propagated to the long-distance matrix elements within the non-relativistic QCD framework~\cite{Butenschoen:2012px}. However, these considerations cannot be directly extended to \Dstar mesons.
In contrast, the polarisation of non-prompt charmonia arises from helicity conservation in the decays of scalar B mesons, analogous to the mechanism expected for \Dstar mesons. The non-prompt \Dstar-meson \rzz measured with respect to the helicity axis is also compared to that of non-prompt \Jpsi and \PsiTwoS mesons in the rapidity interval $|y|<1.2$ measured by the CMS Collaboration in pp collisions at $\s=13~\TeV$~\cite{CMS:2024igk} in Fig.~\ref{fig-withJpsi}. 
Measurements of quarkonium polarisation are traditionally reported in terms of the \lt parameter, which is extracted from the two-dimensional angular distribution of the two muons emerging from their decays. The \lt parameter is related to the \rzz parameter via
\begin{equation}
   \rzz = \frac{1-\lt}{3+\lt},
\end{equation}
implying $\rzz>1/3$ if $\lt<0$ and vice versa.
Given that non-prompt \Dstar, \Jpsi and \PsiTwoS are all vector mesons mainly originating from scalar b-hadron decays, qualitatively similar values of \rzz are obtained. However, a larger \rzz is observed for the \Dstar meson in the overlap \pt region of $20<\pt<30~\GeV/c$. This behaviour is described by PYTHIA 8 + EvtGen simulations and originates from different decays of beauty hadrons into \Dstar, \Jpsi, or \PsiTwoS mesons.

Overall, the results presented in this paper confirm the conclusions drawn from the measurement performed at $\s=13~\TeV$~\cite{ALICE:2022byg} while substantially extending the accessible \pt range and reference frames. They establish a precise baseline for current and more precise future measurements of \Dstar spin alignment in heavy-ion collisions~\cite{ALICE:2025cdf} and provide valuable input to models incorporating spin-dependent effects in heavy-quark fragmentation.

\section{Summary}

In this paper, the measurement of the spin alignment of prompt and non-prompt \Dstar mesons with respect to the helicity and production axes in pp collisions at $\s=13.6~\TeV$ is presented. The significantly increased statistical precision with respect to previous measurements allowed the study to be extended to transverse momenta from $\pt=3~\GeVc$ up to $\pt=100~\GeVc$ ($30~\GeVc$) for prompt (non-prompt) \Dstar mesons. The measured values of the spin-density matrix element \rzz for prompt \Dstar mesons are consistent within uncertainties with the unpolarised expectation of $1/3$ over the full explored \pt range for both quantisation axes. These results are in agreement with previous measurements performed in pp collisions at $\s=13~\TeV$, with earlier observations in \ee collisions, as well as with models based on spin-dependent fragmentation functions, indicating that possible charm-quark polarisation effects are either absent at production or effectively diluted during the hadronisation process in pp collisions.

For non-prompt \Dstar mesons, the measured \rzz values are significantly higher (lower) than $1/3$ in the helicity (production) frames, consistent with expectations from helicity conservation in \HbtoDstar weak decays. These results are qualitatively similar to those reported by the CMS Collaboration for non-prompt \Jpsi and \PsiTwoS mesons in pp collisions at $\s=13~\TeV$.
The results are compatible with predictions obtained from simulations based on PYTHIA 8 coupled with the EvtGen decay package. This observation confirms the presence of spin alignment for non-prompt \Dstar mesons with respect to the helicity and production axes and validates the modelling of spin effects in beauty-hadron decays.

The measurements presented in this paper establish a robust reference for current and future studies of \Dstar spin alignment in heavy-ion collisions. In particular, they provide a well-constrained baseline for disentangling possible medium-induced effects, such as those arising from the large initial angular momentum and magnetic field in nucleus–nucleus collisions and from contributions related to decay kinematics and hadronisation in pp collisions.

%% file: fa_2026-08-11_Opt_C.tex

The ALICE Collaboration would like to thank all its engineers and technicians for their invaluable contributions to the construction of the experiment and the CERN accelerator teams for the outstanding performance of the LHC complex.
The ALICE Collaboration gratefully acknowledges the resources and support provided by all Grid centres and the Worldwide LHC Computing Grid (WLCG) collaboration.
The ALICE Collaboration acknowledges the following funding agencies for their support in building and running the ALICE detector:
A. I. Alikhanyan National Science Laboratory (Yerevan Physics Institute) Foundation (ANSL), State Committee of Science and World Federation of Scientists (WFS), Armenia;
Austrian Academy of Sciences, Austrian Science Fund (FWF): [M 2467-N36] and Nationalstiftung f\"{u}r Forschung, Technologie und Entwicklung, Austria;
Ministry of Communications and High Technologies, National Nuclear Research Center, Azerbaijan;
Rede Nacional de Física de Altas Energias (Renafae), Financiadora de Estudos e Projetos (Finep), Funda\c{c}\~{a}o de Amparo \`{a} Pesquisa do Estado de S\~{a}o Paulo (FAPESP) and The Sao Paulo Research Foundation  (FAPESP), Brazil;
Bulgarian Ministry of Education and Science, within the National Roadmap for Research Infrastructures 2020-2027 (object CERN), Bulgaria;
Ministry of Education of China (MOEC) , Ministry of Science \& Technology of China (MSTC) and National Natural Science Foundation of China (NSFC), China;
Ministry of Science and Education and Croatian Science Foundation, Croatia;
Centro de Aplicaciones Tecnol\'{o}gicas y Desarrollo Nuclear (CEADEN), Cubaenerg\'{\i}a, Cuba;
Ministry of Education, Youth and Sports of the Czech Republic, Czech Republic;
The Danish Council for Independent Research | Natural Sciences, the VILLUM FONDEN and Danish National Research Foundation (DNRF), Denmark;
Helsinki Institute of Physics (HIP), Finland;
Commissariat \`{a} l'Energie Atomique (CEA) and Institut National de Physique Nucl\'{e}aire et de Physique des Particules (IN2P3) and Centre National de la Recherche Scientifique (CNRS), France;
Bundesministerium f\"{u}r Forschung, Technologie und Raumfahrt (BMFTR) and GSI Helmholtzzentrum f\"{u}r Schwerionenforschung GmbH, Germany;
National Research, Development and Innovation Office, Hungary;
Department of Atomic Energy Government of India (DAE), Department of Science and Technology, Government of India (DST), University Grants Commission, Government of India (UGC) and Council of Scientific and Industrial Research (CSIR), India;
National Research and Innovation Agency - BRIN, Indonesia;
Istituto Nazionale di Fisica Nucleare (INFN), Italy;
Japanese Ministry of Education, Culture, Sports, Science and Technology (MEXT) and Japan Society for the Promotion of Science (JSPS) KAKENHI, Japan;
Consejo Nacional de Ciencia (CONACYT) y Tecnolog\'{i}a, through Fondo de Cooperaci\'{o}n Internacional en Ciencia y Tecnolog\'{i}a (FONCICYT) and Direcci\'{o}n General de Asuntos del Personal Academico (DGAPA), Mexico;
Nederlandse Organisatie voor Wetenschappelijk Onderzoek (NWO), Netherlands;
The Research Council of Norway, Norway;
Pontificia Universidad Cat\'{o}lica del Per\'{u}, Peru;
Ministry of Science and Higher Education, National Science Centre and WUT ID-UB, Poland;
Korea Institute of Science and Technology Information and National Research Foundation of Korea (NRF), Republic of Korea;
Ministry of Education and Scientific Research, Institute of Atomic Physics, Ministry of Research and Innovation and Institute of Atomic Physics and Universitatea Nationala de Stiinta si Tehnologie Politehnica Bucuresti, Romania;
Ministerstvo skolstva, vyskumu, vyvoja a mladeze SR, Slovakia;
National Research Foundation of South Africa, South Africa;
Swedish Research Council (VR) and Knut \& Alice Wallenberg Foundation (KAW), Sweden;
European Organization for Nuclear Research, Switzerland;
Suranaree University of Technology (SUT), National Science and Technology Development Agency (NSTDA) and National Science, Research and Innovation Fund (NSRF via PMU-B B05F650021), Thailand;
Turkish Energy, Nuclear and Mineral Research Agency (TENMAK), Turkey;
National Academy of  Sciences of Ukraine, Ukraine;
Science and Technology Facilities Council (STFC), United Kingdom;
National Science Foundation of the United States of America (NSF) and United States Department of Energy, Office of Nuclear Physics (DOE NP), United States of America.
In addition, individual groups or members have received support from:
Czech Science Foundation (grant no. 23-07499S), Czech Republic;
FORTE project, reg.\ no.\ CZ.02.01.01/00/22\_008/0004632, Czech Republic, co-funded by the European Union, Czech Republic;
European Research Council (grant no. 101220549), European Union;
Deutsche Forschungs Gemeinschaft (DFG, German Research Foundation) ``Neutrinos and Dark Matter in Astro- and Particle Physics'' (grant no. SFB 1258), Germany;
CONVECS project, CUP C97H23001700002 FESR 2021-2027 program, Italy.

%% file: Alice_Authorlist_2026-08-11_Opt_C.tex
\begin{flushleft} 
\small

D.A.H.~Abdallah\,\orcidlink{0000-0003-4768-2718}\,$^{\rm 134}$, 
I.J.~Abualrob\,\orcidlink{0009-0005-3519-5631}\,$^{\rm 111}$, 
S.~Acharya\,\orcidlink{0000-0002-9213-5329}\,$^{\rm 49}$, 
A.B.~Adiguzel$^{\rm 91}$, 
K.~Agarwal\,\orcidlink{0000-0001-5781-3393}\,$^{\rm II,}$$^{\rm 23}$, 
G.~Aglieri Rinella\,\orcidlink{0000-0002-9611-3696}\,$^{\rm 32}$, 
L.~Aglietta\,\orcidlink{0009-0003-0763-6802}\,$^{\rm 24}$, 
N.~Agrawal\,\orcidlink{0000-0003-0348-9836}\,$^{\rm 25}$, 
Z.~Ahammed\,\orcidlink{0000-0001-5241-7412}\,$^{\rm 132}$, 
S.~Ahmad\,\orcidlink{0000-0003-0497-5705}\,$^{\rm 15}$
% ...

\section*{Affiliation Notes}

$^{\rm I}$ Deceased\\
$^{\rm II}$ Also at: INFN Trieste, Trieste, Italy\\
$^{\rm III}$ Also at: Fondazione Bruno Kessler (FBK), Trento, Italy\\
$^{\rm IV}$ Also at: Czech Technical University in Prague, Prague, Czech Republic\\
$^{\rm V}$ Also at: Dipartimento DET del Politecnico di Torino, Turin, Italy\\
$^{\rm VI}$ Also at: University College of Dublin, Dublin, Ireland\\
$^{\rm VII}$ Also at: Department of Applied Physics, Aligarh Muslim University, Aligarh, India\\
$^{\rm VIII}$ Also at: Institute of Theoretical Physics, University of Wroclaw, Wroclaw, Poland\\
$^{\rm IX}$ Also at: Facultad de Ciencias, Universidad Nacional Aut\'{o}noma de M\'{e}xico, Mexico City, Mexico\\
$^{\rm X}$ Also at: Lovely Professional University, Punjab, India
\\

\section*{Collaboration Institutes}

$^{1}$ A.I. Alikhanyan National Science Laboratory (Yerevan Physics Institute) Foundation, Yerevan, Armenia\\
$^{2}$ AGH University of Krakow, Cracow, Poland\\
$^{3}$ Bogolyubov Institute for Theoretical Physics, National Academy of Sciences of Ukraine, Kyiv, Ukraine\\
$^{4}$ Bose Institute, Department of Physics  and Centre for Astroparticle Physics and Space Science (CAPSS), Kolkata, India\\
$^{5}$ California Polytechnic State University, San Luis Obispo, California, United States\\
$^{6}$ Central China Normal University, Wuhan, China\\
$^{7}$ Centro de Aplicaciones Tecnol\'{o}gicas y Desarrollo Nuclear (CEADEN), Havana, Cuba\\
$^{8}$ Centro de Investigaci\'{o}n y de Estudios Avanzados (CINVESTAV), Mexico City and M\'{e}rida, Mexico\\
$^{9}$ Chicago State University, Chicago, Illinois, United States\\
$^{10}$ China Nuclear Data Center, China Institute of Atomic Energy, Beijing, China\\
$^{11}$ China University of Geosciences, Wuhan, China\\
$^{12}$ Chungbuk National University, Cheongju, Republic of Korea\\
$^{13}$ Comenius University Bratislava, Faculty of Mathematics, Physics and Informatics, Bratislava, Slovak Republic\\
$^{14}$ Creighton University, Omaha, Nebraska, United States\\
$^{15}$ Department of Physics, Aligarh Muslim University, Aligarh, India\\
$^{16}$ Department of Physics, Pusan National University, Pusan, Republic of Korea\\
$^{17}$ Department of Physics, Sejong University, Seoul, Republic of Korea\\
$^{18}$ Department of Physics, University of California, Berkeley, California, United States\\
$^{19}$ Department of Physics, University of Oslo, Oslo, Norway\\
$^{20}$ Department of Physics and Technology, University of Bergen, Bergen, Norway\\
$^{21}$ Dipartimento di Fisica, Universit\`{a} di Pavia, Pavia, Italy\\
$^{22}$ Dipartimento di Fisica dell'Universit\`{a} and Sezione INFN, Cagliari, Italy\\
$^{23}$ Dipartimento di Fisica dell'Universit\`{a} and Sezione INFN, Trieste, Italy\\
$^{24}$ Dipartimento di Fisica dell'Universit\`{a} and Sezione INFN, Turin, Italy\\
$^{25}$ Dipartimento di Fisica e Astronomia dell'Universit\`{a} and Sezione INFN, Bologna, Italy\\
$^{26}$ Dipartimento di Fisica e Astronomia dell'Universit\`{a} and Sezione INFN, Catania, Italy\\
$^{27}$ Dipartimento di Fisica e Astronomia dell'Universit\`{a} and Sezione INFN, Padova, Italy\\
$^{28}$ Dipartimento di Fisica `E.R.~Caianiello' dell'Universit\`{a} and Gruppo Collegato INFN, Salerno, Italy\\
$^{29}$ Dipartimento DISAT del Politecnico and Sezione INFN, Turin, Italy\\
$^{30}$ Dipartimento di Scienze MIFT, Universit\`{a} di Messina, Messina, Italy\\
$^{31}$ Dipartimento Interateneo di Fisica `M.~Merlin' and Sezione INFN, Bari, Italy\\
$^{32}$ European Organization for Nuclear Research (CERN), Geneva, Switzerland\\
$^{33}$ Faculty of Electrical Engineering, Mechanical Engineering and Naval Architecture, University of Split, Split, Croatia\\
$^{34}$ Faculty of Nuclear Sciences and Physical Engineering, Czech Technical University in Prague, Prague, Czech Republic\\
$^{35}$ Faculty of Physics, Sofia University, Sofia, Bulgaria\\
$^{36}$ Faculty of Science, P.J.~\v{S}af\'{a}rik University, Ko\v{s}ice, Slovak Republic\\
$^{37}$ Faculty of Technology, Environmental and Social Sciences, Bergen, Norway\\
$^{38}$ Frankfurt Institute for Advanced Studies, Johann Wolfgang Goethe-Universit\"{a}t Frankfurt, Frankfurt, Germany\\
$^{39}$ Fudan University, Shanghai, China\\
$^{40}$ Gauhati University, Department of Physics, Guwahati, India\\
$^{41}$ Helmholtz-Institut f\"{u}r Strahlen- und Kernphysik, Rheinische Friedrich-Wilhelms-Universit\"{a}t Bonn, Bonn, Germany\\
$^{42}$ Helsinki Institute of Physics (HIP), Helsinki, Finland\\
$^{43}$ High Energy Physics Group,  Universidad Aut\'{o}noma de Puebla, Puebla, Mexico\\
$^{44}$ Horia Hulubei National Institute of Physics and Nuclear Engineering, Bucharest, Romania\\
$^{45}$ HUN-REN Wigner Research Centre for Physics, Budapest, Hungary\\
$^{46}$ Indian Institute of Technology Bombay (IIT), Mumbai, India\\
$^{47}$ Indian Institute of Technology Indore, Indore, India\\
$^{48}$ INFN, Laboratori Nazionali di Frascati, Frascati, Italy\\
$^{49}$ INFN, Sezione di Bari, Bari, Italy\\
$^{50}$ INFN, Sezione di Bologna, Bologna, Italy\\
$^{51}$ INFN, Sezione di Cagliari, Cagliari, Italy\\
$^{52}$ INFN, Sezione di Catania, Catania, Italy\\
$^{53}$ INFN, Sezione di Padova, Padova, Italy\\
$^{54}$ INFN, Sezione di Pavia, Pavia, Italy\\
$^{55}$ INFN, Sezione di Torino, Turin, Italy\\
$^{56}$ INFN, Sezione di Trieste, Trieste, Italy\\
$^{57}$ Inha University, Incheon, Republic of Korea\\
$^{58}$ Institute for Gravitational and Subatomic Physics (GRASP), Utrecht University/Nikhef, Utrecht, Netherlands\\
$^{59}$ Institute of Experimental Physics, Slovak Academy of Sciences, Ko\v{s}ice, Slovak Republic\\
$^{60}$ Institute of Physics, Homi Bhabha National Institute, Bhubaneswar, India\\
$^{61}$ Institute of Physics of the Czech Academy of Sciences, Prague, Czech Republic\\
$^{62}$ Institute of Space Science (ISS), Bucharest, Romania\\
$^{63}$ Institut f\"{u}r Kernphysik, Johann Wolfgang Goethe-Universit\"{a}t Frankfurt, Frankfurt, Germany\\
$^{64}$ Instituto de Ciencias Nucleares, Universidad Nacional Aut\'{o}noma de M\'{e}xico, Mexico City, Mexico\\
$^{65}$ Instituto de F\'{i}sica, Universidade Federal do Rio Grande do Sul (UFRGS), Porto Alegre, Brazil\\
$^{66}$ Instituto de F\'{\i}sica, Universidad Nacional Aut\'{o}noma de M\'{e}xico, Mexico City, Mexico\\
$^{67}$ iThemba LABS, National Research Foundation, Somerset West, South Africa\\
$^{68}$ Jeonbuk National University, Jeonju, Republic of Korea\\
$^{69}$ Korea Institute of Science and Technology Information, Daejeon, Republic of Korea\\
$^{70}$ Laboratoire de Physique Subatomique et de Cosmologie, Universit\'{e} Grenoble-Alpes, CNRS-IN2P3, Grenoble, France\\
$^{71}$ Lawrence Berkeley National Laboratory, Berkeley, California, United States\\
$^{72}$ Lund University Department of Physics, Division of Particle Physics, Lund, Sweden\\
$^{73}$ Marietta Blau Institute, Vienna, Austria\\
$^{74}$ Nagasaki Institute of Applied Science, Nagasaki, Japan\\
$^{75}$ Nara Women{'}s University (NWU), Nara, Japan\\
$^{76}$ National Centre for Nuclear Research, Warsaw, Poland\\
$^{77}$ National Institute of Science Education and Research, Homi Bhabha National Institute, Jatni, India\\
$^{78}$ National Nuclear Research Center, Baku, Azerbaijan\\
$^{79}$ National Research and Innovation Agency - BRIN, Jakarta, Indonesia\\
$^{80}$ Niels Bohr Institute, University of Copenhagen, Copenhagen, Denmark\\
$^{81}$ Nikhef, National institute for subatomic physics, Amsterdam, Netherlands\\
$^{82}$ Nuclear Physics Group, STFC Daresbury Laboratory, Daresbury, United Kingdom\\
$^{83}$ Nuclear Physics Institute of the Czech Academy of Sciences, Husinec-\v{R}e\v{z}, Czech Republic\\
$^{84}$ Oak Ridge National Laboratory, Oak Ridge, Tennessee, United States\\
$^{85}$ Ohio State University, Columbus, Ohio, United States\\
$^{86}$ Physics Department, Panjab University, Chandigarh, India\\
$^{87}$ Physics Department, University of Jammu, Jammu, India\\
$^{88}$ Physics Program and International Institute for Sustainability with Knotted Chiral Meta Matter (WPI-SKCM$^{2}$), Hiroshima University, Hiroshima, Japan\\
$^{89}$ Physikalisches Institut, Eberhard-Karls-Universit\"{a}t T\"{u}bingen, T\"{u}bingen, Germany\\
$^{90}$ Physikalisches Institut, Ruprecht-Karls-Universit\"{a}t Heidelberg, Heidelberg, Germany\\
$^{91}$ Physik Department, Technische Universit\"{a}t M\"{u}nchen, Munich, Germany\\
$^{92}$ Politecnico di Bari and Sezione INFN, Bari, Italy\\
$^{93}$ Research Division and ExtreMe Matter Institute EMMI, GSI Helmholtzzentrum f\"ur Schwerionenforschung GmbH, Darmstadt, Germany\\
$^{94}$ Saga University, Saga, Japan\\
$^{95}$ Saha Institute of Nuclear Physics, Homi Bhabha National Institute, Kolkata, India\\
$^{96}$ School of Physics and Astronomy, University of Birmingham, Birmingham, United Kingdom\\
$^{97}$ Secci\'{o}n F\'{\i}sica, Departamento de Ciencias, Pontificia Universidad Cat\'{o}lica del Per\'{u}, Lima, Peru\\
$^{98}$ SUBATECH, IMT Atlantique, Nantes Universit\'{e}, CNRS-IN2P3, Nantes, France\\
$^{99}$ Sungkyunkwan University, Suwon City, Republic of Korea\\
$^{100}$ Suranaree University of Technology, Nakhon Ratchasima, Thailand\\
$^{101}$ Technical University of Ko\v{s}ice, Ko\v{s}ice, Slovak Republic\\
$^{102}$ The Henryk Niewodniczanski Institute of Nuclear Physics, Polish Academy of Sciences, Cracow, Poland\\
$^{103}$ The University of Texas at Austin, Austin, Texas, United States\\
$^{104}$ Universidad Aut\'{o}noma de Sinaloa, Culiac\'{a}n, Mexico\\
$^{105}$ Universidade de S\~{a}o Paulo (USP), S\~{a}o Paulo, Brazil\\
$^{106}$ Universidade Estadual de Campinas (UNICAMP), Campinas, Brazil\\
$^{107}$ Universidade Federal do ABC, Santo Andre, Brazil\\
$^{108}$ Universitatea Nationala de Stiinta si Tehnologie Politehnica Bucuresti, Bucharest, Romania\\
$^{109}$ University of Cape Town, Cape Town, South Africa\\
$^{110}$ University of Derby, Derby, United Kingdom\\
$^{111}$ University of Houston, Houston, Texas, United States\\
$^{112}$ University of Jyv\"{a}skyl\"{a}, Jyv\"{a}skyl\"{a}, Finland\\
$^{113}$ University of Kansas, Lawrence, Kansas, United States\\
$^{114}$ University of Liverpool, Liverpool, United Kingdom\\
$^{115}$ University of Science and Technology of China, Hefei, China\\
$^{116}$ University of Silesia in Katowice, Katowice, Poland\\
$^{117}$ University of South-Eastern Norway, Kongsberg, Norway\\
$^{118}$ University of Tennessee, Knoxville, Tennessee, United States\\
$^{119}$ University of the Witwatersrand, Johannesburg, South Africa\\
$^{120}$ University of Tokyo, Tokyo, Japan\\
$^{121}$ University of Tsukuba, Tsukuba, Japan\\
$^{122}$ University of Zagreb Faculty of Science, Department of Physics, Zagreb, Croatia\\
$^{123}$ Universit\"{a}t M\"{u}nster, Institut f\"{u}r Kernphysik, M\"{u}nster, Germany\\
$^{124}$ Universit\'{e} Clermont Auvergne, CNRS/IN2P3, LPC, Clermont-Ferrand, France\\
$^{125}$ Universit\'{e} de Lyon, CNRS/IN2P3, Institut de Physique des 2 Infinis de Lyon, Lyon, France\\
$^{126}$ Universit\'{e} de Strasbourg, CNRS, IPHC UMR 7178, F-67000 Strasbourg, France, Strasbourg, France\\
$^{127}$ Universit\'{e} Paris-Saclay, Centre d'Etudes de Saclay (CEA), IRFU, D\'{e}partment de Physique Nucl\'{e}aire (DPhN), Saclay, France\\
$^{128}$ Universit\'{e}  Paris-Saclay, CNRS/IN2P3, IJCLab, Orsay, France\\
$^{129}$ Universit\`{a} degli Studi di Foggia, Foggia, Italy\\
$^{130}$ Universit\`{a} del Piemonte Orientale, Vercelli, Italy\\
$^{131}$ Universit\`{a} di Brescia, Brescia, Italy\\
$^{132}$ Variable Energy Cyclotron Centre, Homi Bhabha National Institute, Kolkata, India\\
$^{133}$ Warsaw University of Technology, Warsaw, Poland\\
$^{134}$ Wayne State University, Detroit, Michigan, United States\\
$^{135}$ Yale University, New Haven, Connecticut, United States\\
$^{136}$ Yildiz Technical University, Istanbul, Turkey\\
$^{137}$ Yonsei University, Seoul, Republic of Korea\\
$^{138}$ Affiliated with an institute formerly covered by a cooperation agreement with CERN\\
$^{139}$ Affiliated with an international laboratory covered by a cooperation agreement with CERN.\\

\end{flushleft} 